\documentclass{aa}
\usepackage{caption}
\usepackage{graphicx}
\usepackage{subfigure}
\usepackage{array}
\usepackage{tabularx}
\usepackage{color}
\usepackage{soul}
\usepackage{natbib}
\usepackage{threeparttable}
\usepackage[colorlinks=true,citecolor=blue]{hyperref}
\setstcolor{red}
\usepackage{txfonts}

\begin{document} 

\title{Atmospheric composition of WASP-85Ab with ESPRESSO/VLT observations}

\titlerunning{Atmospheric composition of WASP-85Ab}

   \author{Zewen Jiang
          \inst{1}\fnmsep\inst{2},
          Wei Wang \inst{1}\thanks{E-mail: wangw@nao.cas.cn}, Guo Chen \inst{3}, Fei Yan \inst{4}, Heather M. Cegla\inst{5,6}\thanks{UKRI Future Leaders Fellow}, Patricio Rojo \inst{7}, Yaqing Shi \inst{1}\fnmsep\inst{2}, Qinlin Ouyang \inst{1}\fnmsep\inst{2}, Meng Zhai\inst{8}, 
          Yujuan Liu\inst{1}, Fei Zhao\inst{1}, Yuqin Chen \inst{1}
          }
   \authorrunning{Zewen Jiang et al.}
   \institute{CAS Key Laboratory of Optical Astronomy, National Astronomical Observatories, Chinese Academy of Sciences, Datun Road A20, Beijing 100101, China
         \and
             School of Astronomy and Space Science, University of Chinese Academy of Sciences, Beijing 100049, China
        \and
            Key Laboratory of Planetary Sciences, Purple Mountain Observatory, Chinese Academy of Sciences, Nanjing 210033, PR China
        \and 
            Department of Astronomy, University of Science and Technology of China, Hefei 230026, China
        \and 
            Physics Department, University of Warwick, Coventry CV4 7AL, United Kingdom
        \and 
            Centre for Exoplanets and Habitability, University of Warwick, Coventry CV4 7AL, UK
        \and
            Departamento de Astronomía, Universidad de Chile, Camino El Observatorio 1515, Las Condes, Santiago, Chile
        \and
            CAS South America Center for Astronomy, National Astronomical Observatories, Chinese Academy of Sciences, Datun Road A20, Beijing 100101, China
             }

   \date{Received xxx ; accepted xxx}

\abstract{Transit spectroscopy is the most frequently used technique to reveal the atmospheric properties of exoplanets, while that at high resolution has the advantage to resolve the small Doppler shift of spectral lines, and the trace signal of the exoplanet atmosphere can be separately extracted. We obtain the transmission spectra of the extrasolar planet WASP-85Ab, a hot Jupiter in a 2.655-day orbit around a G5, $V=11.2$ mag host star, observed by high-resolution spectrograph ESPRESSO at the Very Large Telescope array for three transits. We present an analysis of the Rossiter-McLaughlin effect on WASP-85A, and determine a spin-orbit angle ${\lambda = -16.155^{\circ}}^{+2.916}_{-2.879}$, suggesting that the planet is in an almost aligned orbit. Combining the transmission spectra of three nights, we tentatively detected H$\alpha$ and \ion{Ca}{II} absorption with $\gtrapprox 3\sigma$ via direct visual inspection of the transmission spectra with the Center-to-Limb variation and the Rossiter-McLaughlin effects removed, which still remain visible after excluding the cores of these strong lines with a 0.1\AA\,mask. These spectral signals seems likely to origin from the planetary atmosphere, but we can not fully exclude their stellar origins. Via the cross-correlation analysis of a set of atoms and molecules, \ion{Li}{I} is marginally detected at $\sim4\sigma$ level, suggesting that Li might be present in the atmosphere of WASP-85Ab. }

   \keywords{planetary Systems -- planets and satellites: individual: WASP-85Ab -- planet and satellites: atmospheres -- methods: observational -- techniques: spectroscopic
               }

   \maketitle
%
\section{Introduction}

Characterization of exoplanets is one of the fastest-growing and most exciting fields in astronomy in recent years. The information encoded in the exoplanet atmosphere can provide critical insights into myriad atmospheric processes as well as the formation and evolutionary history of the planet~\citep{Madhusudhan_2019}. Thanks to their relatively high equilibrium temperatures and large atmosphere scale heights, hot Jupiters (HJs) have relatively prominent atmosphere signals. This makes HJs the best targets to carry out detailed characterization of the physical and chemical properties of exoplanet atmospheres. Among all the powerful tools for the atmospheric characterization of exoplanets, high-resolution spectroscopy (HRS) becomes widely applied to search for atoms and molecules whose transmission spectra have either several strong individual lines or a few dense forests of spectral lines. HRS is sensitive to the change of exoplanet radial velocity, since the host star and Earth moves much slower than a planet, so that the atmospheric signals of an exoplanet are separated from the stellar and telluric signals. For the same reason, HRS can be used to characterize the planet’s atmosphere due to its sensitivity to the depth, shape, and position of the planet’s spectral lines, Compared to low-resolution spectroscopy, HRS has the advantage to probe above cloud decks where the cores of the strongest spectral lines are formed~\citep{birkby2018exoplanet}.

The first ground-based detection of CO was achieved by HRS in the hot Jupiters HD\,209458b  \citep{Snellen_2010}. Plenty of atoms and molecules have been detected in the atmospheres of many HJs including Na, K, Li, Ca, Cr, Mg, He, Ti, Fe, Fe+, and Ca+ as well as H$_2$O, etc. For example, CH4 and Na are detected in the atmospheres of HD\,209458b~\citep{Giacobbe_2021,Charbonneau_2002}, Cr in WASP-189b \citep{Prinoth_2022}, K in WASP-121b \citep{Merritt_2021}, Sc and Ti+ in HD\,189733b \citep{Lira_Barria_2021}, Ti and TiO in HD\,149026b \citep{Ishizuka_2021}. All these species are detected by the HRS method.

The planet WASP-85Ab~\citep{Brown_2014} was identified in both the Super-WASP and WASP-South photometry independently and confirmed by spectroscopic follow-up using the SOPHIE spectrograph~ \citep{Perruchot_2008} mounted on the 1.93-m telescope of the Observatoire de Haute Provence. It is a classic HJ with a mass of $1.265\pm0.065$\,$M_{\rm Jup}$, a radius of $1.240\pm0.030$\,$R_{\rm Jup}$ and an equilibrium temperature $T_{\rm eq}$ of 1452\,K with an orbital period of $\sim$2.65\,d~\citep{Mocnik_2016}. Thus, it is an interesting target for atmospheric characterisation given its moderately high equilibrium temperature and the low density. Up to now, no such study has been performed yet. WASP-85Ab orbits a G5V star in a binary star system. The host star WASP-85A is active with starspots detected by \citet{Mocnik_2016} from their analysis on the K2 short-cadence data of WASP-85A. 


   \begin{table*}
      \caption[]{Summary of the WASP-85Ab transit observations}
         \label{obs}
         \begin{tabular}{lccccccccc}
            \hline
            \hline
            \noalign{\smallskip}
            & Date&VLT/UT &\multicolumn{3}{c}{Number of spectra} & Exp. time & Airmass range & Mean S/N & Program ID\\
            \cline{4-6}
            \noalign{\smallskip}
            & (UT Time) &  &Total & In-transit & Out-of-transit & (s) &  & (@550\,nm) &   \\
            \noalign{\smallskip}
            \hline
            \noalign{\smallskip}
            Night 1 & 2021-02-13 &UT3 & 47 & 21 & 26 &400 & $1.1-1.7$ & $\sim$64 & 0106.D-0853  \\
            \noalign{\smallskip}
            Night 2 & 2021-02-21 &UT4& 41 & 22 & 19 &400 & $1.1-1.6$ & $\sim$63 & 0106.D-0853 \\
            \noalign{\smallskip}
            Night 3 & 2021-03-17 &UT2& 39 & 21 & 18 &400& $1.1-1.5$ & $\sim$61 & 0106.D-0853   \\  
            \noalign{\smallskip}
            \hline
         \end{tabular}
         \newline
         
     \end{table*}
%

This paper is organized as follows. We present the details of three transit observations of WASP-85Ab using the Echelle Spectrograph for Rocky Exoplanet and Stable Spectroscopic Observations (ESPRESSO) in Section~\ref{sec:obs and reduction}. We measure the orbital architecture of the WASP-85A system analyzed based on the Rossiter-McLaughlin (RM) effects in Section~\ref{sec:data_analysis}. The high-resolution transmission spectra are presented in Section~\ref{sec:transmission spectrum analysis}, followed by our cross-correlation analysis of the spectra. The discussion and conclusions are presented in Section~\ref{sec:dis and conclusion}.

\section{Observations and data reduction}
\label{sec:obs and reduction}
Three transits of WASP-85A were observed in February and March 2021 with ESPRESSO~\citep{Pepe_2021} under the ESO programs 0106.D-0853~(PI: H.~M.~Cegla). They aimed to detect and characterize stellar differential rotation, and convection-induced variations and to determine spin-orbit alignment, which can help us to better understand the properties of WASP-85A (H.M Cegla et al. \textit{in prep}). While in this study, we will mainly target to reveal the composition of WASP-85Ab atmosphere. 

ESREESSO is a fibre-fed ultra-stable échelle high-resolution spectrograph, mounted at the 8.2m Very Large Telescope at  European Southern Observatory in  Cerro Paranal, Chile~\citep{Pepe_2021}. The observations were taken in the High Resolution 1-UT mode, and data were read out in $2\times1$ binning and slow readout mode, which yields a spectral resolving power $R$ of $\sim$140\,000 and a wavelength range of 380$\sim$788\,nm. During the observations, fiber A was pointed to the target star, while fiber B was pointed to the sky for simultaneous monitoring of the sky emission. 

The three transits were observed on the following dates: 2021 Feb. 13 (Night 1, the first transit, hereafter T1), Feb. 21 (Night 2, the second transit, T2), and Mar. 17 (Night 3, the third transit, T3), each covering an entire transit plus 1 to 2 hours out-of-transit baselines. Individual exposure time for each transit is 400\,s. In T1, a total of 47 spectra were obtained with 21 in transit and 26 out of transit covering the orbital phase $\Phi$ from $-0.062$ to $+0.031$. In T2, 41 spectra were acquired including 22 in-transit and 19 out-of-transit spectra with $\Phi$ from $-0.043$ to $+0.037$. In T3, the number of exposures collected in total is 39 with $\Phi$ ranging from $-0.036$ to $+0.044$. The numbers of the in- and out-of-transit spectra are 21 and 18, respectively. Details of the three observations are summarized in  Table~\ref{obs}. 

WASP-85A has a K0-dwarf stellar companion namely WASP-85B with a $V$-mag difference of 0.7\,mag and an angular distance of $\sim1.5$\arcsec~\citep{Brown_2014}, larger than the ESPRESSO fiber size of 1\arcsec. Assuming a 2D Gaussian PSF with seeing sizes of 1\arcsec\ and 0.85\arcsec, we determine that WASP-85B's flux contamination to 85A is less than 0.44\% and 0.14\%. This suggests that at least for the T2 and T3 data, which were mostly taken with seeing better than 0.85\arcsec, should be mostly clean from WASB-85B's contamination.

\begin{figure}
   \centering
   \includegraphics[width=8cm, height=12cm]{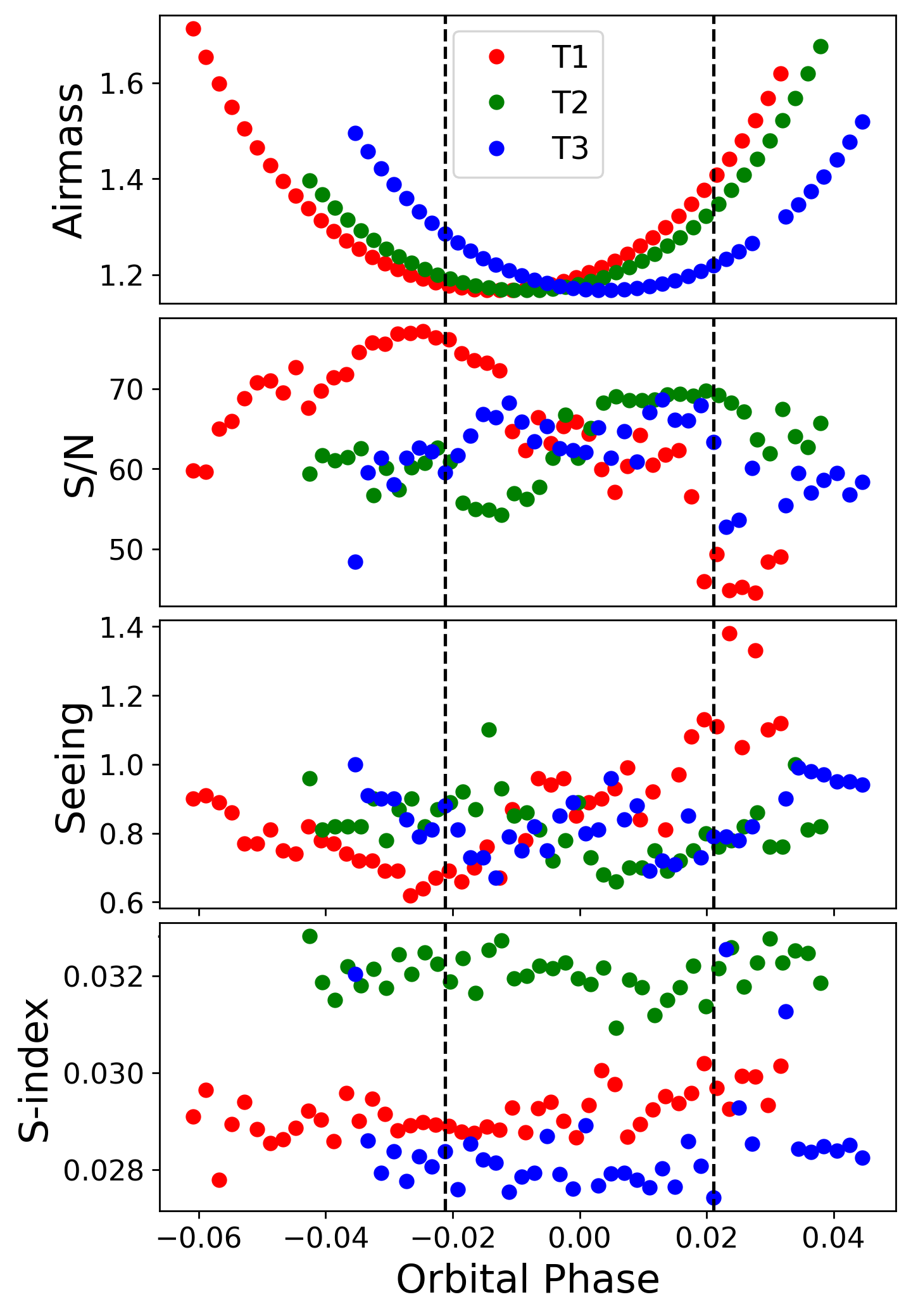}
 \caption{The temporal variations of the airmass, S/N@550\,nm, seeing and S-index of the three transit observations from top to bottom, respectively. The green dashed lines indicate the first and fourth contacts of the transits.}
    \label{fig:airmass_snr}
    \end{figure}

The temporal variations of the airmass, S/N ratio, seeing and S-index during the three observing nights are shown in Fig.~\ref{fig:airmass_snr}. The stellar activity level of WASP-85A as represented by the S-index is the lowest in T3 and the highest in T2, suggesting that the data taken in T3 should be the least affected by stellar activity while those in T2 are affected most. It is also worthy to note that the median and variance of the seeing size in T1 are larger than those in T2 and T3, with quite a number exceeding 1\arcsec. Given that the fiber size of ESPRESSO is 1\arcsec\ and the distance between the WASP-85A and B is $\sim1.5$\arcsec~\citep{Brown_2014}, at least some of the spectra taken in T1 may bear both slit-loss and light contamination from the stellar companion, which will lead to lower S/N ratios and weaker planet signals. On the other hand, most of T2 \& T3 data points should have only minor slit loss and should not be affected by the companion's contamination. Combining the above discussion, and the fact that planet signal is the strongest or only observable in T3, we use the T3 data as the main data set in this work.

For data analysis, we used the one-dimensional sky-corrected spectra processed by the ESPRESSO reduction pipeline, version 2.3.4, downloaded from the ESO advanced science archive. The raw spectral images were corrected for bias, dark, flat, and bad pixels before being extracted into 1D spectra. 

   \begin{table}
      \caption[]{Molecfit parameters used for telluric correction}
         \label{molecfit parameter}
                  \begin{tabular}{lll}
            \hline
            \hline
            \noalign{\smallskip}
             Parameter & Value & Description \\
            \noalign{\smallskip}
            \hline
            \noalign{\smallskip}
           ftol & $10^{-9}$& ${\chi}^2$\ tolerance \\
            \noalign{\smallskip}
           xtol & $10^{-9}$& Tolerance for the \texttt{molecfit}   \\
                &          & fitted variables   \\
            \noalign{\smallskip}
           fit\textunderscore cont & 1 & Continuum fitting flag  \\ 
            \noalign{\smallskip}
            cont\textunderscore n & 3 & Degree of polynomial continuum \\
            \noalign{\smallskip}
           fit\textunderscore res\textunderscore gauss & 1 &  Gaussian kernel  \\
            \noalign{\smallskip}
            res\textunderscore gauss& 3.5 & Kernel size (pixels) \\  
            \noalign{\smallskip}
            kernfac & 6.0 & Kernel size measured in units of \\
                    &     & the kernel FWHM \\
            \noalign{\smallskip}
            list\textunderscore molec & H$_2$\textnormal{O}, O$_2$ & Molecules to be synthetised   \\
            \noalign{\smallskip}
            \noalign{\smallskip}
            \hline
         \end{tabular}
      
   \end{table}

   \begin{table}
      \caption[]{Physical and orbit parameters of the WASP-85 system}
         \label{system} 
           \begin{tabularx}{0.5\textwidth}{lll}
            \hline
            \hline
            \noalign{\smallskip}
            Description &Symbol & Value \\
            \noalign{\smallskip}
            \hline
            \noalign{\smallskip}
            Stellar Parameters & & \\
            \noalign{\smallskip}
            \hline
            \noalign{\smallskip}
            $V$ magnitude& $m_{\rm v}$    & $11.2\pm0.011$\,mag    \\
            \noalign{\smallskip}       
            Effective temp& $T_{\rm eff}$& $6112\pm27$\,K   \\          
            \noalign{\smallskip}
            Surface gravity&  log\,$g_\star$  & $4.48\pm0.11$ cgs\\    
            \noalign{\smallskip}
            Metallicity& [Fe/H] &  $0.00\pm0.05$\,dex   \\
            \noalign{\smallskip}
             Stellar mass&  $M_\star$  & $1.09\pm0.08\,M_\odot$   \\
            \noalign{\smallskip}    
             Stellar radius &$R_\star$ & $0.935\pm0.0023\,R_\odot$   \\
            \noalign{\smallskip} 
            Right ascension &  R. A. & $11^{\rm h}43^{\rm m}38.01^{\rm s}$ \\                  
            \noalign{\smallskip}
            Declination &  Dec & $+06^{\circ}33^{\prime}49.4^{\prime\prime}$  \\
            \hline
            \noalign{\smallskip}
		Planet Parameter & & \\
            \noalign{\smallskip}  
             \hline
            \noalign{\smallskip}
            Planet mass   & $M_{\rm p}$ & $1.265\pm0.065\,M_{\rm Jup}$   \\        
           \noalign{\smallskip}
            Planet radius & $R_{\rm p}$ & $1.24\pm0.03\,R_{\rm Jup}$ \\           
           \noalign{\smallskip}
           Planet density &$\rho$       & $0.660\pm0.020$\,g\,cm$^{-3}$ \\          
            \noalign{\smallskip}
           Eequilibrium temp& $T_{\rm eq}$ & $1452\pm6$\,K  \\     
             \noalign{\smallskip}
            Radius ratio& $R_{\rm p}/R_\star$ & $0.0187\pm0.00002$\\   
             \noalign{\smallskip}
             \hline
             \noalign{\smallskip}
            Orbit Parameters&&\\
            \noalign{\smallskip}  
             \hline
             \noalign{\smallskip}
             Epoch$-2450000$& $T_{\rm c}$ & $6847.472856\pm0.000014$\,BJD \\ 
             \noalign{\smallskip}
              Semi-amplitude$^{1}$&$K_\star$ & $173.3\pm 1.8$\,m\,s$^{-1}$\,\\ 
             \noalign{\smallskip}
             Period & $P$  & $2.6556777\pm0.00000044$\,d    \\ 
             \noalign{\smallskip}
             Transit duration& $T_{14}$ & $0.10817\pm0.00002$\,d \\ 
             \noalign{\smallskip}
             Ingress duration& $T_{12}$ & $0.013037\pm0.000002$\,d \\ 
             \noalign{\smallskip}
             Semi-major axis & $a$ &  $0.039\pm0.001$\,AU   \\ 
            \noalign{\smallskip}
            Inclination &$i$ & $89.69^{+0.11}_{-0.03}$\,deg \\   
            \noalign{\smallskip}
            \hline
         \end{tabularx}
         \tablebib{ \citet{Mocnik_2016},\,\citet{Brown_2014}$^{1}$}
   \end{table}
 
\section{Data analysis}
\label{sec:data_analysis}
\subsection{Telluric correction}
\label{subsec:tel_correction}
The telluric correction was firstly conducted by subtracting the fiber B on-sky spectra from the fiber A on-target spectra for telluric emission features, which had been performed by the ESPRESSO reduction pipeline. We then corrected the telluric absorption imprinted on the obtained spectra using the ESO software \texttt{Molecfit} version 1.5.7~\citep{Smette_2015}, following \citet{Allart_2017}. 

\texttt{Molecfit} is based on synthetic modeling of the Earth’s atmospheric transmission with a line-by-line radiative transfer model, which has been widely used recently in high-resolution transmission spectroscopic studies. \citet{Allart_2017} removed telluric features in the spectrum using Molecfit in order to search for water vapor of HD 189733b; \citet{Hoeijmakers_2019} detected absorption of \ion{Na}{I}, \ion{Cr}{II}, \ion{Sc}{II} and \ion{Y}{II} in the atmosphere of an ultra-hot Jupiter, KELT-9b, after removing telluric contamination; \citet{Tabernero_2021} modeled the telluric transmission spectrum using  Molecfit to find the absorption of  \ion{Li}{I}, \ion{Na}{I}, \ion{Mg}{I}, \ion{Ca}{II}, \ion{K}{I} and \ion{Fe}{I}. 

Fig.~\ref{fig:Telluric_correction_effect} presents comparisons of a sample original spectrum and its corresponding spectrum with telluric correction applied around the Na D1 $\&$ D2-line doublet region (Top panel) and H$\alpha$ spectral line region (Bottom panel) on the night of 2018 October 31, which illustrate that the telluric features have been corrected quite well. The specific parameters used in this work for telluric correction are listed in Table~\ref{molecfit parameter}. Note that the obtained ESPRESSO spectra are given in the solar system barycentric rest frame, while \texttt{Molecfit} requires terrestrial rest frame spectra as input. Therefore, we shifted the observed spectra to the terrestrial rest frame considering the barycentric Earth radial velocity (BERV) for a better correction of telluric absorption.

\subsection{Rossiter-McLaughlin analysis}
\label{subsec:RME}
When a planet transits its rotating host star, the portion of the stellar disk blocked by the planet varies with time, resulting in anomalous radial-velocity variation that will overlay on the Doppler reflex motion. This effect was first pointed out by \citet{Rossiter_1924} and \citet{McLaughlin_1924}, and is known as the RM effect~\citep{Triaud_2018}. \citet{Queloz_2000} reported the first discovery of the RM effect on an exoplanet, by analyzing the anomaly in the radial velocity curve after removing the influence of Doppler reflex motion. From this anomaly, they determined $\upsilon sin \textit{i}_{\star}$ and the angle between the orbital plane and the apparent equatorial plane, namely the spin-orbit angle $\lambda$.

Two main methods are widely used to measure the radial velocities (RVs), one is based on least-squares fitting by carrying out the template matching of the observed spectra \citep{Butler_1996}; the other  utilizes the cross-correlations function (CCF) of observation spectra with the binary mask \citep{Baranne_1996,Pepe_2002}. In this work, the RVs and their uncertainties determined using the CCF method by the ESPRESSO pipeline are adopted.

To analyze and model the RM effect based on the RVs, we use Markov chain Monte Carlo (MCMC) algorithm which is available in \texttt{emcee}~\citep{Foreman-Mackey_2013} and employ the RM effect model presented in \citet{Ohta_2005} implemented in PyAstronomy \citep{Czesla_2019} as \texttt{modelSuite.RmcL}, which performs the analytical fit for the RM effect of a circular orbit. The \texttt{Rmcl} model incorporates with 9 parameters, including the planet's orbital period $P$, the time of transit midpoint $T_{\rm c}$, the inclination of planetary orbit plane $i$, the inclination of the stellar rotation axis $i_\star$, the ratio of planetary and stellar radius $R_{\rm p}/R_\star$, the semi-major axis $a$, the linear limb-darkening coefficient $\epsilon$, the stellar angular rotation velocity $\Omega$, the sky-projected angle $\lambda$ which is the angle between the stellar rotation axis and the normal of planetary orbit plane. Before modeling the RM effect, we estimated and removed the corrected noise in the observed RVs with Gaussian process (GP)~\citep{Foreman_Mackey_2017}, and corrected for the Doppler reflex motion using Kepler's Laws. The rest RVs are used as input for the RM effect model. Note that there are four RV points in T3 deviating from the RV trend which are excluded from the RM analysis. 

Assuming a circular planetary orbit, the RV curve with the baseline trend removed can be expressed with the following formula:
\begin{equation} \label{srv}
	V_\star=K_\star\,{\rm sin}(2\pi \phi)+V_{\rm bary}+V_{\rm sys}
\end{equation}

where $V_\star$ is the RV of the host star, $K_\star$ is the RV semi-amplitude, $\phi$ is the orbital phase of the planet, $V_{\rm bary}$ is the barycentric velocity because of the Earth's revolution, $V_{\rm sys}$ is the radial velocity of targeted exoplanet system. Among them, $V_\star$, $\phi$ and $V_{\rm bary}$ are functions of time, while the values of $K_\star$ and $V_{\rm sys}$ for each night are derived from our GP analysis as listed in Table~\ref{K_Vsys}. Here, we use the $K_\star$ value shown in Table~\ref{system} and fixed it in the next steps.

\begin{figure}
   \centering
   \includegraphics[width=8cm, height=8cm]{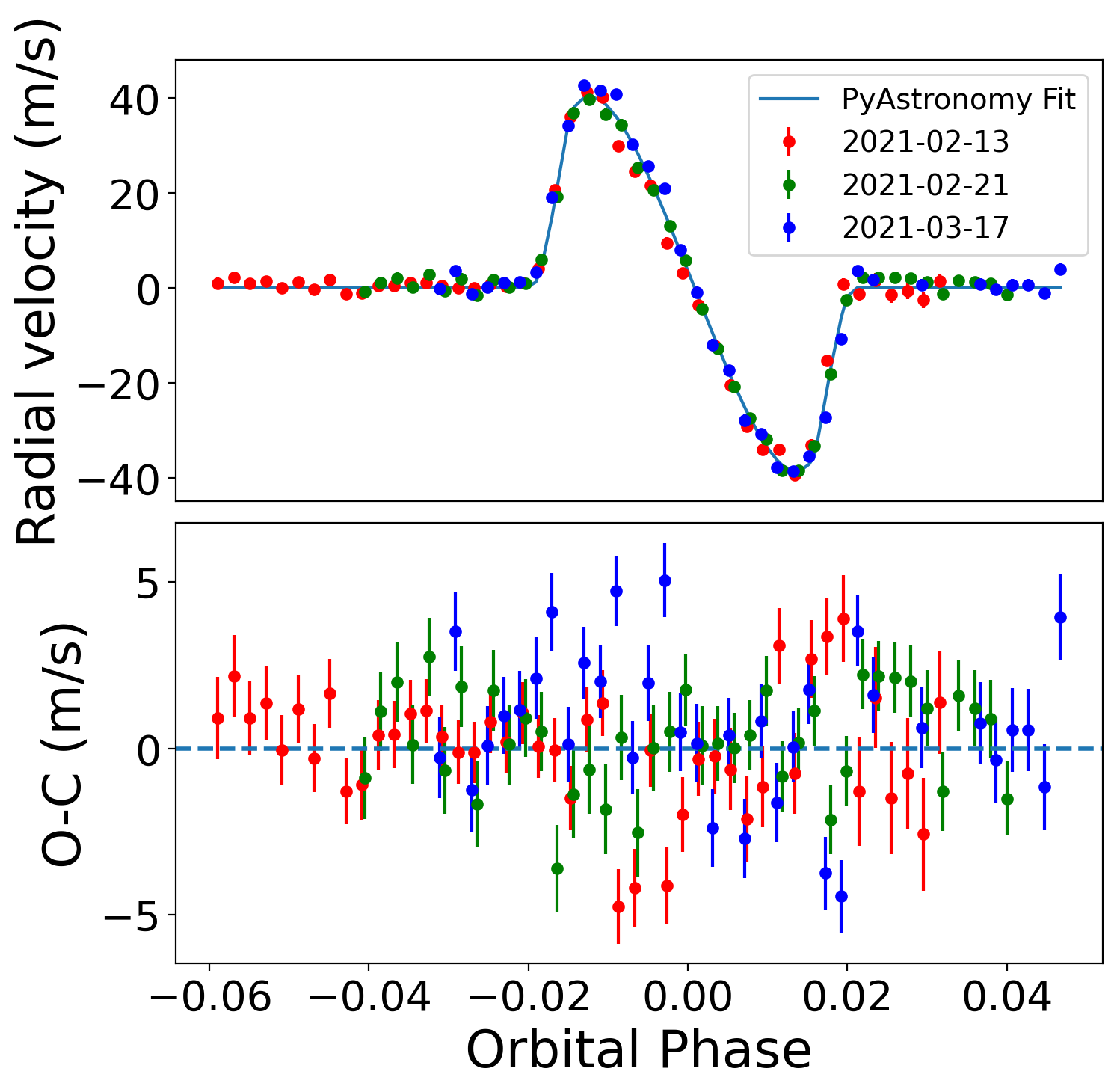}
 \caption{The RM effect fit using the GP+PyAstronomy package on the ESPRESSO RVs taken from the spectrum headers. \emph{Top panel:} The RV curve with Kepler motion and red noise removed, showing the RM effect of WASP-85Ab of the three transits represented by the red, green, and blue circles with error bars. The best-fit model using the PyAstronomy package is shown in the solid line. \emph{Bottom panel:} The residuals between the data and the model predictions.}
    \label{fig:clv+RM fit}
    \end{figure}

We set $\lambda$, $\epsilon$, $\Omega$, and $i_{\star}$ as free parameters for our model, with their priors listed in Table~\ref{prior}. The remaining parameters including $a$, $P$, $i$, $\textit{T}_c$ and $R_{p}/R_{\star}$ are fixed to the specific values as listed in Table~\ref{system}. In order to explore a wide parameter space, we set the walkers to 200 with 50,\,000 steps. The first 10,\,000 steps were discarded as burn-in in order to probe the parameter space and get settled into the maximum density in the MCMC chain. The thus obtained parameters from the MCMC fitting are shown in Table~\ref{prior}, whereas the retrieved best model is compared with the data set in Fig.~\ref{fig:clv+RM fit}, with an rms residual of $\sim 1.877$\,m s$^{-1}$. The derived projected spin-orbit angle $\lambda$ = -16.155$^{\circ}$$^{+2.916}_{-2.879}$, implying the planetary orbit plane is roughly aligned to the stellar equator. We also derive the projected stellar rotation velocity $v{\rm sin}\,i_{\star} = 2.98^{+0.050}_{-0.043}$ $\rm km\,s^{-1}$, the linear limb dark coefficient $\epsilon$ =$0.855^{+0.014}_{-0.014}$. These values are used as input to model the center-to-limb variation (CLV) effect and RM effect in order to eliminate the influence caused by the occultation of the host star by the planet. The derived posterior distributions for the three transits are given in Fig.~\ref{fig:corner} in detail.

   \begin{table*}
      \caption[]{Parameters derived from the RV curve fitting.}
         \label{prior}
         $$
         \begin{tabular}{p{0.28\linewidth}cccc}
            \hline
            \hline
            \noalign{\smallskip}
             Description &Symbol& Prior & Posterior   \\
            \noalign{\smallskip}
            \hline
            \noalign{\smallskip}
           Projected spin-orbit angle&$\lambda$ &U(-50,\,50)&$-16.155^{+2.916}_{-2.879}$\,deg& \\
            \noalign{\smallskip}
           Linear limb dark coefficient&$\epsilon$ & U(0.5,\,1)&$0.855^{+0.014}_{-0.014}$\\
            \noalign{\smallskip}
           Projected stellar rotation velocity&$\Omega$ &U($3\rm e^{-6}$,\,$1\rm e^{-5}$)&$0.00000468^{+1.57\rm e^{-7}}_{-8.59\rm e^{-8}}$ rad\,s$^{-1}$ \\  
            \noalign{\smallskip}
           Inclination of stellar rotation axis&$i_\star$ & U(70,\,110) & $91.6947^{+12.7688}_{-14.7692}$\,deg \\  
            \noalign{\smallskip}
            \hline
         \end{tabular}
         $$
         \textbf{Notes}:\textit{U(a,b)} represents a uniform distribution with a low and high limit of a and b, respectively.
   \end{table*}

\begin{table}
\caption[]{Parameters derived from the GP analysis and the estimated noise of the three transit observations.}
\label{K_Vsys}
\begin{tabular}{llll}
 \hline
 \hline
 \noalign{\smallskip}
Parameter & T1 & T2 & T3   \\
\noalign{\smallskip}
\hline
\noalign{\smallskip}
$K_\star$~(m\,s$^{-1}$) & $170^{+24}_{-15}$&$157^{+4}_{-0}$&$165^{+20}_{-12}$\\
\noalign{\smallskip}
$V_{\rm sys}$~(km\,s$^{-1}$) &$13.531^{+0.004}_{-0.005}$&$13.530^{+0.001}_{-0.001}$&$13.519^{+0.024}_{-0.005}$\\
\noalign{\smallskip}
GP noise med. & 1.000105 & 0.999998  &1.000064\\
\noalign{\smallskip}
GP noise std.  &  0.000367 &  0.000087 &  0.000115 \\
\noalign{\smallskip}
\hline
\end{tabular}
\end{table}

\section{Transmission spectrum analysis}
\label{sec:transmission spectrum analysis}
The atmosphere composition of HJs plays an important role in understanding atmospheric dynamics and planetary evolution. Here we mainly search for the most abundant atoms including  \ion{Na}{I}, \ion{Mg}{I}, H$\alpha$, H$\beta$, \ion{Ca}{II}\,H, \ion{Ca}{II}\,K and \ion{Li}{I}. We adopt the method outlined in \citet{Wyttenbach_2015} and \citet{Casasayas_Barris_2019} to extract transmission spectra of individual lines. As described in Section~\ref{sec:data_analysis}, the observed spectra have been corrected for telluric contamination. Then the spectra are normalized to their corresponding continuum level with a two-degree spline curve fit with the regions around strong absorption lines excluded. We also apply a sigma-clipping rejection algorithm on the normalized spectra and replace the cosmic ray hits with the mean value of all the other spectra at each wavelength~\citep{Allart_2017}.

In order to align the stellar lines, we first shift the spectra to the stellar rest frame, taking into account BERV, $V_{\rm sys}$ and the stellar reflex motion induced by the planet $V_{\rm reflex}$. BERV is obtained directly from the files header information of the ESPRESSO spectra, while $V_{\rm sys}$ is  derived from the RM analysis as described in Section~\ref{subsec:RME} and $V_{\rm reflex}$ is calculated using the orbital parameters in Table~\ref{system}. Then the aligned out-of-transit spectra are averaged with their mean signal-to-noise ratio (SNR) as their weights to construct a master out-of-transit spectrum, which is used to divide each individual spectrum in order to remove stellar light contribution so that the absorption signal from the planetary atmosphere may be exposed~\citep{Stangret_2021}. Next, we shifted each divided in-transit spectrum to the planet rest frame using the planet radial velocity $V_{\rm p}$ determined from $K_{\rm p}$ and the orbit phase of a given time with the following formula:
\begin{equation} \label{prv}
	\Re(\lambda) = \sum_{in}{\frac{F_{in}(\lambda)}{F_{out}}}\bigg|_\textnormal{{Planet\  RV\ Shift}}\      -1
\end{equation}
Finally, the transmission spectrum is the SNR-weighted sum of in-transit residuals.

\begin{figure} 
   \includegraphics[width=9cm, height=5.5cm]{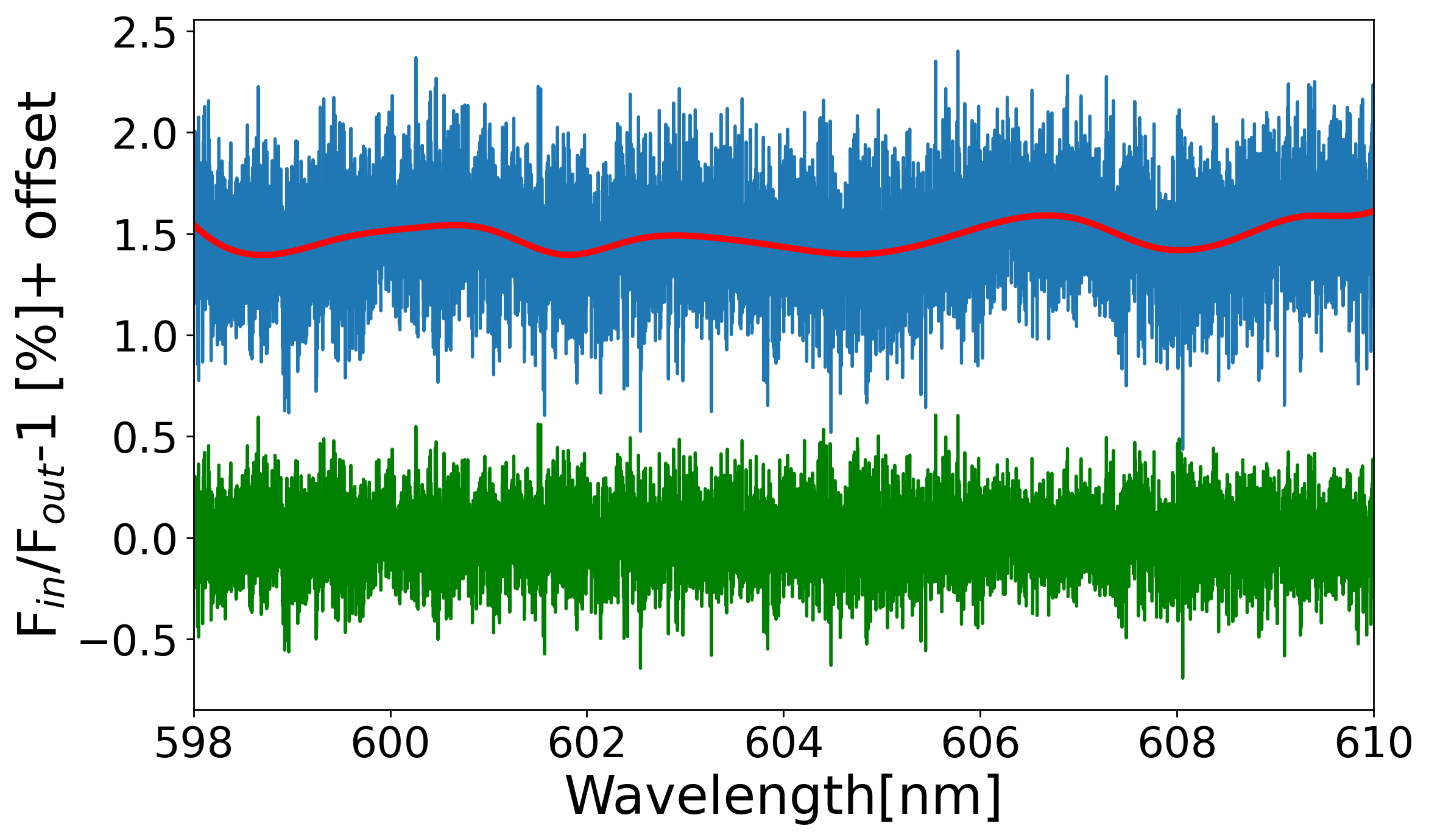}
   \centering
   \caption{An example showing the wiggle pattern in one transmission spectrum taken in T2 in blue, the fitted model in red, and the wiggle-corrected spectrum in green.}
   \label{wiggle}
    \end{figure}

Note that there is a wiggle pattern in the derived transmission spectrum, with an amplitude of $\sim$0.075\% and a period of $\sim$30\,\AA, which might potentially affect the final spectral signature analysis and need to be corrected.\,These wiggles were first reported in \citet{Allart_2020} and are possibly induced by an interference pattern caused by the Coudé train optics as pointed out by \citet{Tabernero_2021}. To correct the wiggles, we remove a sinusoidal trend by fitting a set of cubic splines to the obtained transmission spectra. Fig.~\ref{wiggle} shows an example of such correction. 

The CLV and RM effect affect stellar lines profile and may produce additional time-correlated signals in the transmission spectra. These two effects are both induced by the joint effect of planet occultation and stellar rotation. We adopt the method as presented in \citet{Yan_2018}, \citet{Chen_2020} and \citet{Casasayas_Barris_2019} to model the stellar spectra at different transit positions, which considers both the RM and CLV effects. We use the \texttt{Spectroscopy Made Easy} tool (SME,\citet{Valenti_1996}) to compute the theoretical stellar spectra at 21 different limb-darkening angles ($\mu$) using the MARCS and VALD3 line lists~\citep{Ryabchikova_2015}. We employ solar abundance and local thermodynamical equilibrium (LTE) for the calculation of stellar spectra. 

The stellar disc is then divided into elements of size $0.01\,R_\star\times0.01\,R_\star$, each element owning specific parameters including $v\,{\rm sin}\, i_{\star}$, $\mu$, and $\theta$, which is the angle between the normal to each element and the line of sight. The position of the planet relative to the stellar disk is calculated assuming a uniform velocity during the transit. Then, the synthetic spectrum during transit is calculated by integrating all the surface elements which are not obscured by the planet \citep{Yan_2018}, considering the obscured elements, the proper RV shift, and the corresponding interpolated $\mu$ spectrum.

\begin{figure} 
   \includegraphics[width=9cm, height=5.5cm]{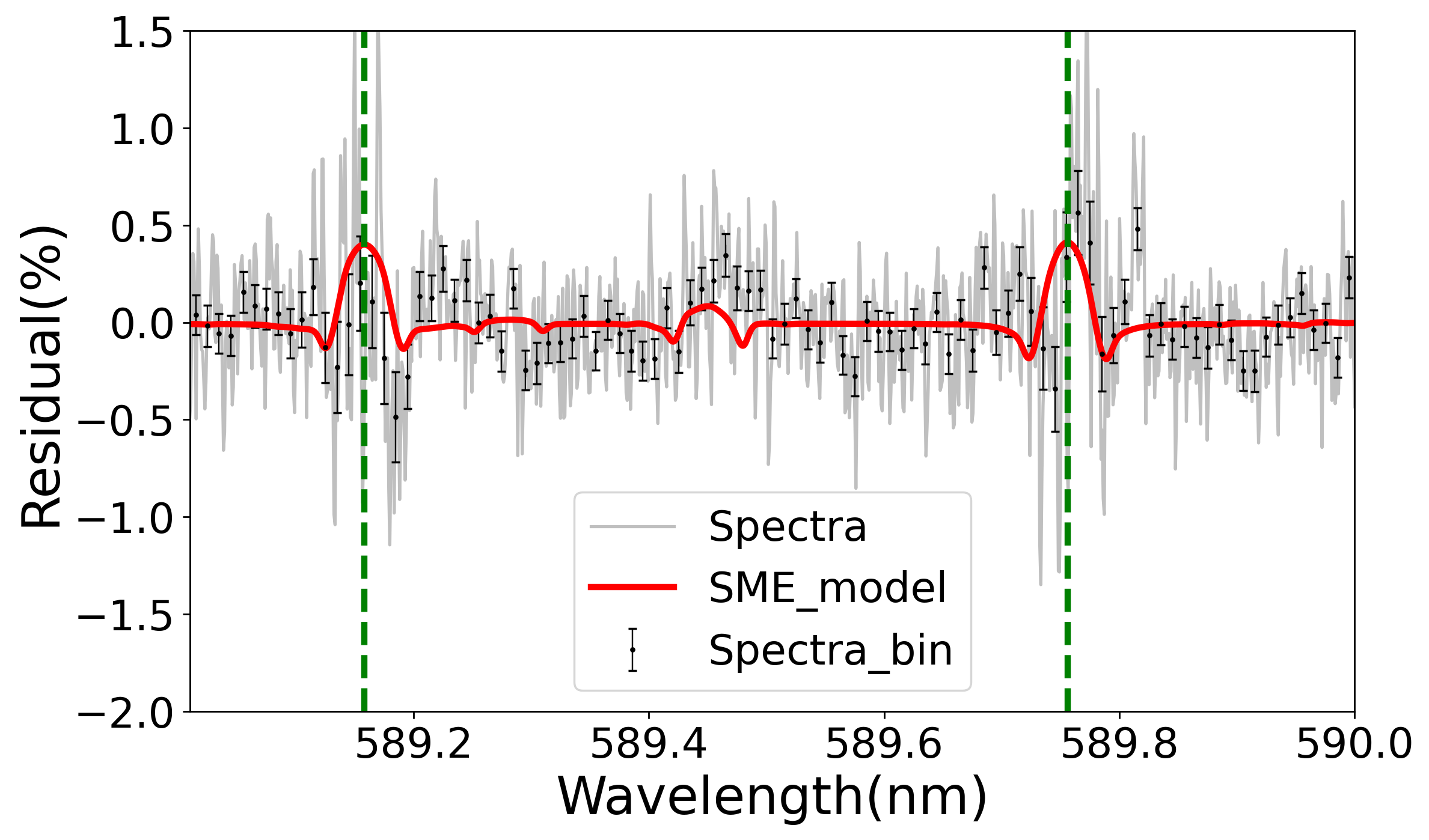}
   \centering
   \caption{The simulated CLV and RM effects overplotted in red on the transmission spectrum. The  transmission spectrum is shown in gray, and the spectrum binned by 0.1{\AA} is shown in black.}
   \label{clvrm}
    \end{figure}

The next step is to simulate the CLV and RM effects and correct them. We construct the master  out-of-transit spectrum of the synthetic spectrum and divided each synthetic spectrum by it, and then we shift each in-transit residual of the synthetic spectrum to the planet rest frame and calculated the sum of all in-transit residuals which only contains the CLV and RM effects. The thus simulated spectra are re-scaled to match the observed spectra and are subtracted to eliminate the influence of CLV and RM effects. Fig.~\ref{clvrm} shows the CLV and RM effects in the residual spectra around the Na D1 and D2 lines.
\subsection{The strong atomic lines}
\label{subsec:stronglines}
\begin{figure*}[htbp]
  \includegraphics[width=\textwidth]{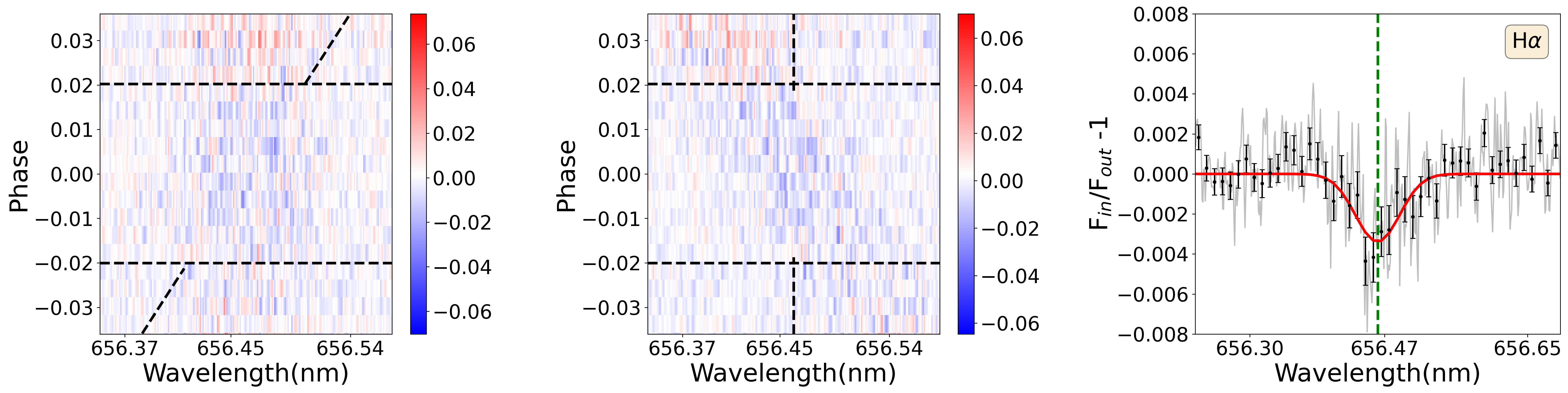}
  \caption{The phase-resolved 2D transmission spectra based on the combination of three nights around the H$\alpha$ line without and with RM+CLV correction applied in the left panel. The horizontal black-dashed lines indicate the beginning and end of the transit. The inclined black-dashed line presents the expected trace of signal from the exoplanet atmosphere. The in-transit part of this line is not plotted so as not to interfere with the visibility of the trace signal. The middle panel shows the 2D transmission spectra in the planet rest frame (PRF) assuming $K_{\rm p}$= 159.76\,km\,s$^{-1}$, and the vertical black-dashed lines indicate the position of expected signal from the exoplanet atmosphere in the planet rest frame. The right panel shows the combined integrated transmission spectrum of H$\alpha$. The grey line represents the transmission spectra in the planet rest frame which has been corrected for the CLV+RM effects, the black dotted line is the binned version with a bin size 0.1 {\AA}, while the best Gaussian fit of the H$\alpha$ line is shown in red. The dashed green vertical line represents the static position of H$\alpha$ at vacuum wavelength. }
  \label{fig:Ha_line}
\end{figure*}

We then performed a visual inspection on the phase-resolved transmission spectra around each known strong feature, which will allow us to confirm or deny the presence of the atoms of interest~\cite {Tabernero_2021}. An integrated SNR-weighted transmission spectrum will be obtained and fitted with a Gaussian function to determine the center, depth, and width of each line. Fig.~\ref{fig:Ha_line} depicts the 2D phase-resolved transmission spectra and the integrated spectrum around H$\alpha$, showing a clear detection of H absorption from the WASP-85Ab's atmosphere. Fig.~\ref{fig:different atom line} shows the same plots for other atomic lines, including the \ion{Na}{I}, \ion{Mg}{I}, \ion{Li}{I}, H$\alpha$, H$\beta$, \ion{Ca}{II} H and K lines. The Gaussian fit results are listed in Table~\ref{atmo_fit_result}. In addition, Table~\ref{atmo_fit_result} also lists the observed Doppler shift of the line center ($V_{wind}$), which traces the planetary winds towards the observer from the morning hemisphere to the evening hemisphere, and effective wavelength-dependent planet radius $R_{\lambda}$, which can be derived with $\sqrt{1+h/\delta}$~$R_p$, where $h$ is the line depth corresponding to different atoms, and $\delta$ is the transit depth of the planet.

As shown in Table~\ref{atmo_fit_result} and Fig.~\ref{fig:Ha_line}, H$\alpha$ absorption is marginally visible in the 2D spectrum (the left panel), and is clearly detected in the integrated spectrum (the right panel), with a depth of $\sim$0.625 $\pm$ 0.06~\%, a full width at half maximum (FWHM) of $\sim23.806\pm2.50$\,km \,s$^{-1}$, a $R_{\lambda}$ of $1.028\pm0.003R_{\rm p}$. Clear absorption is detected for the \ion{Ca}{II} doublet in the integrated transmission spectra (cf. the two bottom panels of Fig.~\ref{fig:different atom line}). FWHM and $R_{\lambda}$ values are similar with those of H$\alpha$. However, it is not visible in the H line 2D spectrum and is only marginally noticeable for the K line. The derived $V_{wind}$ for the H \& K lines are $\sim -4.805$ and 5.974\,km\,s$^{-1}$, respectively, while the obvious inconsistency may arise from the T1 and T2 spectra that suffer large uncertainties due to stellar activity and/or companion's flux contamination. As shown in the top two rows of Fig~\ref{fig:atom line-2}, while the the H \& K lines in T3 are both close to the expected line centers, they are obviously deviated from the line centers -- either blue-shifted or red-shifted. There seems to be 3$\sigma$ detection for \ion{Li}{I} and no clear detection for the other species including \ion{Na}{I}, \ion{Mg}{I}, and H$\beta$. For those with poor Gaussian fits, i.e., \ion{Mg}{I} and H$\beta$, upper limits of the absorption depths are estimated and listed in Table~~\ref{atmo_fit_result}. As shown in Fig.~\ref{fig:atom line-1} and Fig.~\ref{fig:atom line-2}, the T3 spectrum provides the most significant signal for all the species, while the T1 spectrum provides the least information. This is reasonable according to the discussion on the observing conditions in Section 2.

Note that the \ion{Na}{I} D2 absorption line in T1 is abnormal, bearing huge error bars as shown in Fig.~\ref{fig:atom line-1}. Through visual inspection of every 2D raw images, we find that the \ion{Na}{I} D2 line in nearly half of the T1 spectra is contaminated by two bright spikes, while the corresponding 1d spectra without sky subtraction show strong emission at 589\,nm. Examples are given in Fig.~\ref{fig:Na_problem}. These spikes appear simultaneously in both the target and sky fibers and some of them are saturated, leading to close-to-zero counts and thus large uncertainties after sky subtraction. Given the fact that these bright spikes only present around \ion{Na}{I} D2 at 589\,nm, it is quite likely that they arise from projected laser signal leaked from the VLT Laser Guide star.

In order to further confirm the existence of these signals, we perform Empirical Monte Carlo (EMC) simulations~\citep{Redfield_2008} to estimate the effect of systematics, which is widely used previously~\citep{Wyttenbach_2015, Casasayas_Barris_2019, Seidel_2019, Allart_2020}. The idea of this method is to artificially create new data set by randomizing original data, to verify whether the investigated signals still exist. We explore three scenarios covering in-in, in-out, out-out for each strong line, with 10000 iterations for the $\sim$1.5\,{\AA}\ passband. The results are shown in Fig.~\ref{fig:emc for atom}. The in-out distributions of \ion{Li}{I}, H${\alpha}$, \ion{Ca}{II} H and \ion{Ca}{II} K exhibit excess absorption, while the absorption depths in the in-in and out-out scenarios center at zero. The EMC simulations suggest that the signals of \ion{Li}{I}, H${\alpha}$, \ion{Ca}{II} H \& K are likely to be created by the transits and may origin from the planet. In addition, we also mask the line cores with a width of 0.1 {\AA}, where the SNRs are low and the RM+CLV residuals may still exist. These results are shown in Fig.~\ref{fig:masked atom line} and the fitted result is listed in Table~\ref{atmo_fit_result}.

\begin{table*}
\begin{threeparttable}
\caption[]{The derived parameters of the atomic lines from the 3-night combined transmission spectrum$^1$}
\label{atmo_fit_result}
       
\begin{tabular}{p{0.2\linewidth}cccccc}
 \hline
 \hline
 \noalign{\smallskip}
 Line & $\lambda$ & h & V$_{wind}$ & FWHM & R$_{\lambda}$\\
      &[nm]&[\%]& [km s$^{-1}$]&[km s$^{-1}$]&[$R_{\rm p}$]&\\
 \hline
 \noalign{\smallskip}
 \ion{Ca}{II} K &393.478&2.283 $\pm$\ 0.389& 8.536 $\pm$\ 2.528& 28.584$\pm$\ 6.144 &1.512$\pm$\ 0.145 & \\
 \noalign{\smallskip}
\ion{Ca}{II} K (masked) &-&1.769 $\pm$\ 0.372& 8.952 $\pm$\ 3.359& 30.705$\pm$\ 8.180 &1.413$\pm$\ 0.148 & \\
 \noalign{\smallskip}
 \ion{Ca}{II} H &396.959&2.410 $\pm$\ 0.476 &$-4.201\pm2.378$ & 23.646 $\pm$\ 5.713& 1.535 $\pm$\ 0.175 & \\
 \noalign{\smallskip}
\ion{Ca}{II} H (masked) &-&1.729 $\pm$\ 0.497& $-3.495\pm3.105$& $21.294\pm7.458$ &1.405$\pm$\ 0.199 & \\
 \noalign{\smallskip}
 \ion{Mg}{I} & 457.238 &$\leq0.017$ $\pm$\ 0.035 & -&- & $\leq1.005$ $\pm$\ 0.01&\\
 \noalign{\smallskip}
 H$\beta$ & 486.271 &$\leq0.050$ $\pm$\ 0.010& -& -&  $\leq1.014$ $\pm$\ 0.006& \\  
 \noalign{\smallskip}
 \ion{Na}{I} D2$^2$ & 589.158 &$\leq0.069$ $\pm$\ 0.073 & - & - & $\leq1.019$ $\pm$\ 0.040 & \\
 \noalign{\smallskip}
 \ion{Na}{I} D1  & 589.756 & $\leq0.067$ $\pm$\ 0.077& - & - &$\leq1.018$ $\pm$\ 0.043 & \\
 \noalign{\smallskip}
 H$\alpha$ & 656.461 &0.341 $\pm$\ 0.120 &$-1.786 \pm 3.098$ & 18.890 $\pm$\ 7.109& 1.091 $\pm$\ 0.060 & \\  
 \noalign{\smallskip}
 H$\alpha$ (masked) & - &0.350 $\pm$\ 0.14 &$-3.025 \pm2.431$ & $12.899\pm5.659$& $1.096\pm 0.072$ & \\
 \noalign{\smallskip}
 \ion{Li}{I} & 670.961 &0.089$\pm$\ 0.033 &4.223$\pm$\ 6.167 &33.951$\pm$\ 14.425 &1.025$\pm$\ 0.018 & \\ 
 \noalign{\smallskip}
 \ion{Li}{I} (masked) & -&$0.080\pm0.031$ &$5.554\pm6.514$ &$38.307\pm15.243$ &$1.025 \pm 0.017$ & \\ 
 \noalign{\smallskip}
 \hline
\end{tabular}
\begin{tablenotes}
\item[1] The line center wavelength is in vacuum, the depth of absorption line $h$, the Doppler shift of line center $V_{\rm wind}$, line width (FWHM) and the effective planetary radius $R_{\lambda}$
\item[2]The \ion{Na}{I} D2 measurement is obtained on the transmission spectra generated from only the T2 and T3 data.
\end{tablenotes}
\end{threeparttable}
\end{table*}
\begin{figure*}[htbp]
    \centering
    \subfigure{
    \begin{minipage}[t]{1\linewidth}
    \centering
    \includegraphics[width=16.5cm]{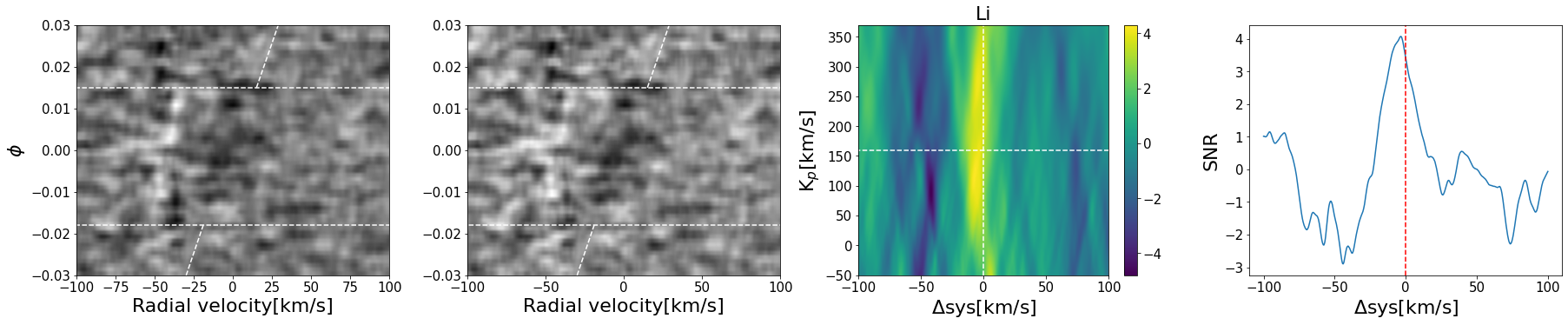}  
    \end{minipage}
    }
    \subfigure{
    \begin{minipage}[t]{1\linewidth}
    \centering
    \includegraphics[width=16.5cm]{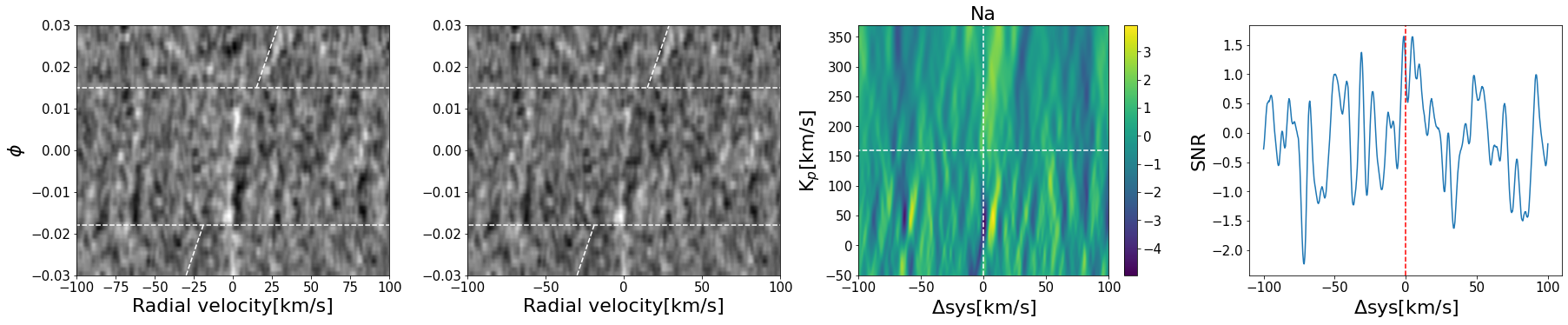}  
    \end{minipage}
    }
    
    \subfigure{
    \begin{minipage}[t]{1\linewidth}
    \centering
    \includegraphics[width=16.5cm]{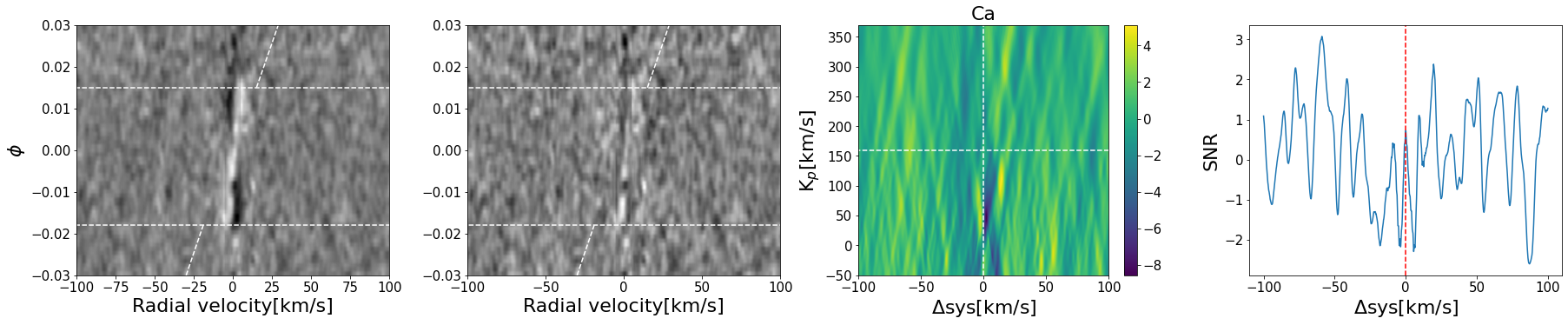}  
    \end{minipage}
    }
    
    
    \subfigure{
    \begin{minipage}[t]{1\linewidth}
    \centering
    \includegraphics[width=16.5cm]{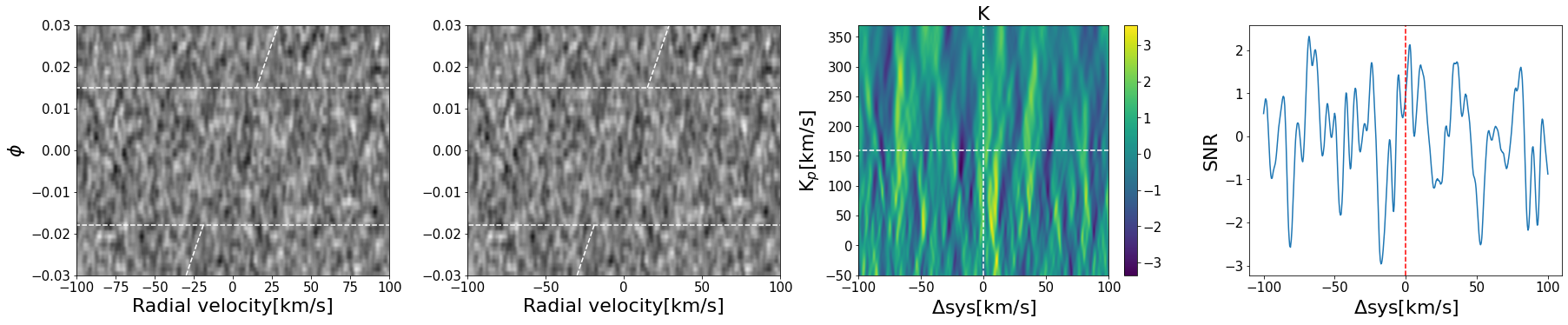}  
    \end{minipage}
    }

    \subfigure{
    \begin{minipage}[t]{1\linewidth}
    \centering
    \includegraphics[width=16.5cm]{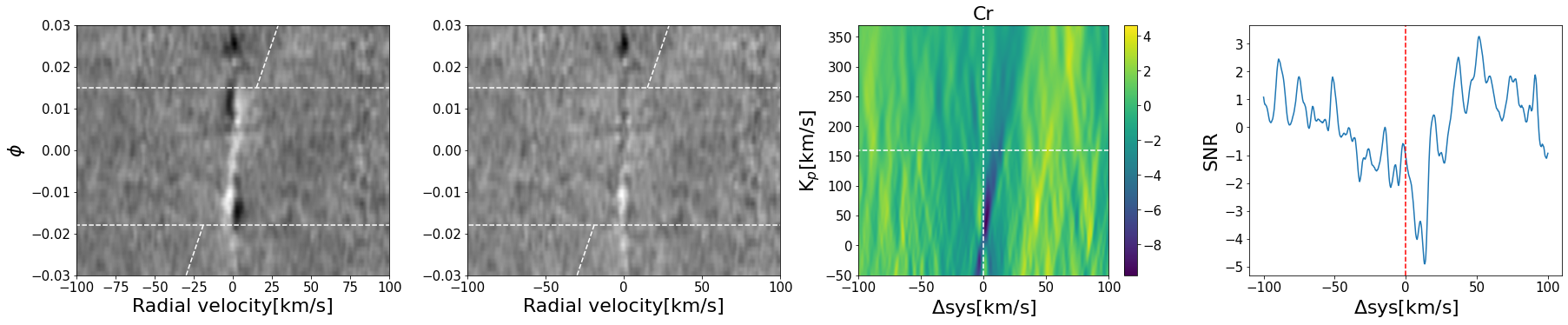}  
    \end{minipage}
    }
    
    \subfigure{
    \begin{minipage}[t]{1\linewidth}
    \centering
    \includegraphics[width=16.5cm]{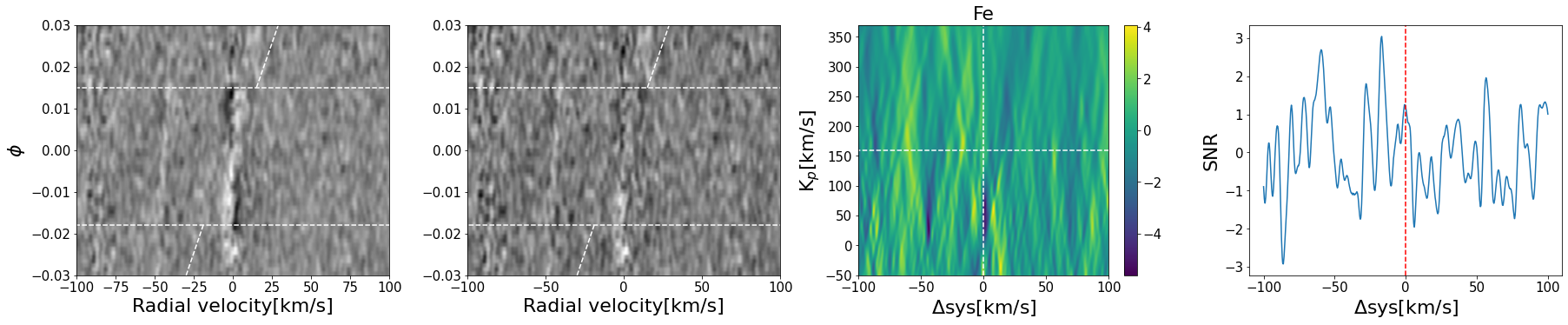}  
    \end{minipage}
    }

\caption{The CCF results derived by cross-correlating the template spectra produced by petitRADTRANS with the observed transmission spectra. \emph{First panels}: The 2D CCF maps of  \ion{Li}{I}, \ion{Na}{I}, \ion{Ca}{I}, \ion{K}{I}, \ion{Cr}{I}, and \ion{Fe}{I} with CLV+RM effects uncorrected . The white dotted lines mark the beginning and ending position of the transit, and the inclined white lines  indicate the expected trace of signal from the planet. \emph{Second panels}: Same as \emph{the first panels} but with CLV+RM effects corrected. \emph{Third panels}: the $K_p$-$\Delta V_{\rm sys}$ maps in the range of $-50\sim350$\, km\,s$^{-1}$. The signal is expected to appear around the intersection of two white dotted lines. \emph{Fourth panels}: the SNR plots at the expected $K_p$ position. }
    \label{ccf-result1}
\end{figure*}
\begin{figure*}[htbp]
    \centering
    \subfigure{
    \begin{minipage}[t]{1\linewidth}
    \centering
    \includegraphics[width=17cm]{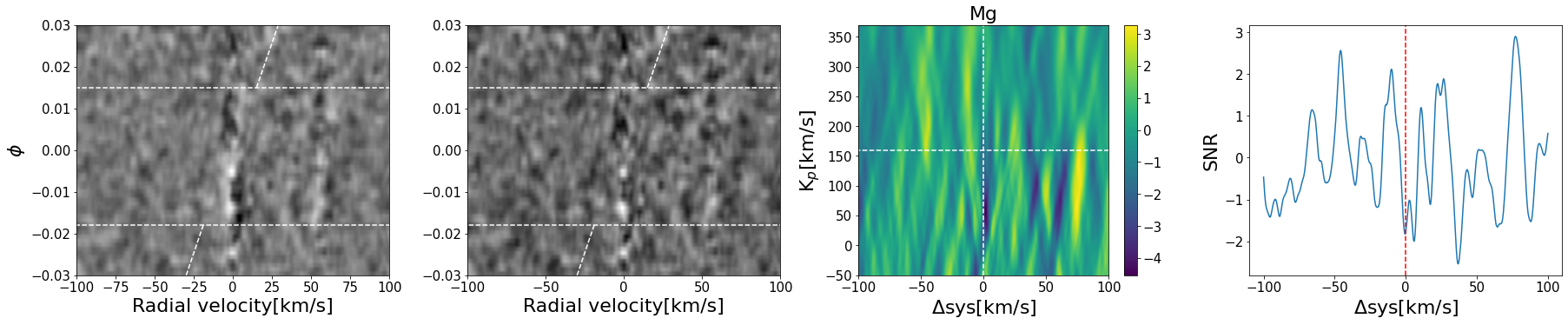}  
    \end{minipage}
    }
    \subfigure{
    \begin{minipage}[t]{1\linewidth}
    \centering
    \includegraphics[width=17cm]{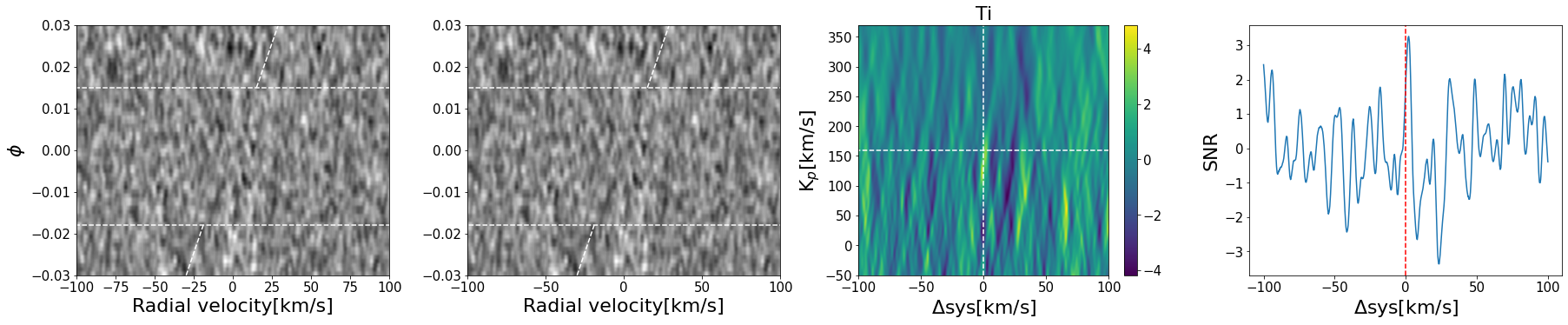}  
    \end{minipage}
    }    
    \subfigure{
    \begin{minipage}[t]{1\linewidth}
    \centering
    \includegraphics[width=17cm]{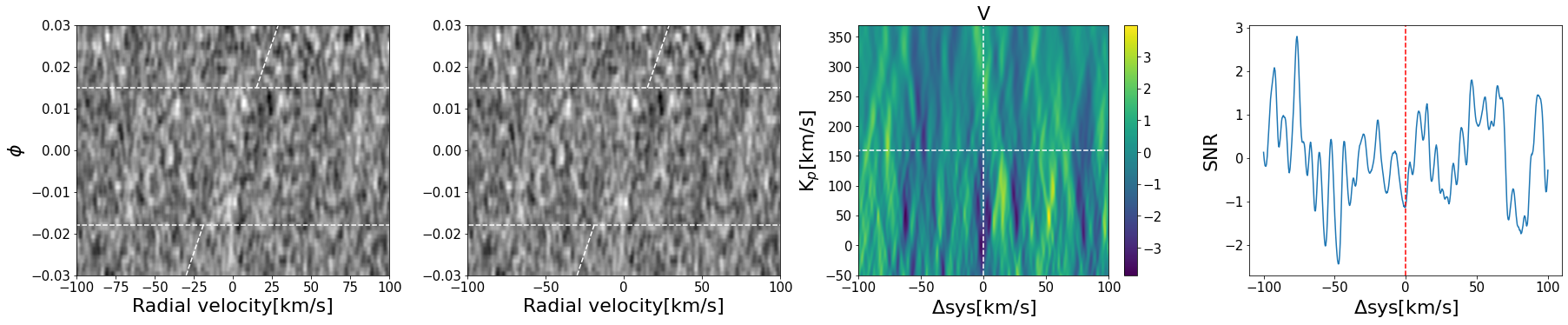}  
    \end{minipage}
    }
    \subfigure{
    \begin{minipage}[t]{1\linewidth}
    \centering
    \includegraphics[width=17cm]{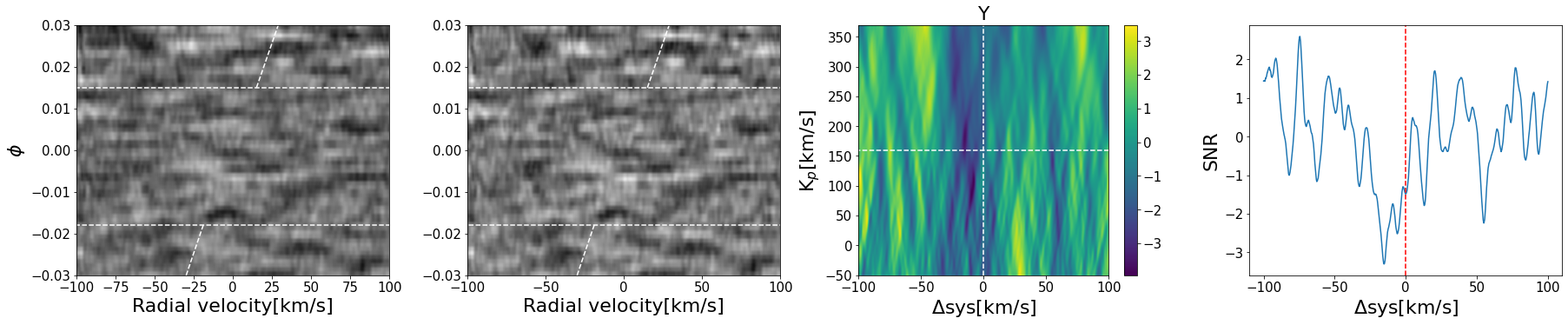}  
    \end{minipage}
    }
    \subfigure{
    \begin{minipage}[t]{1\linewidth}
    \centering
    \includegraphics[width=17cm]{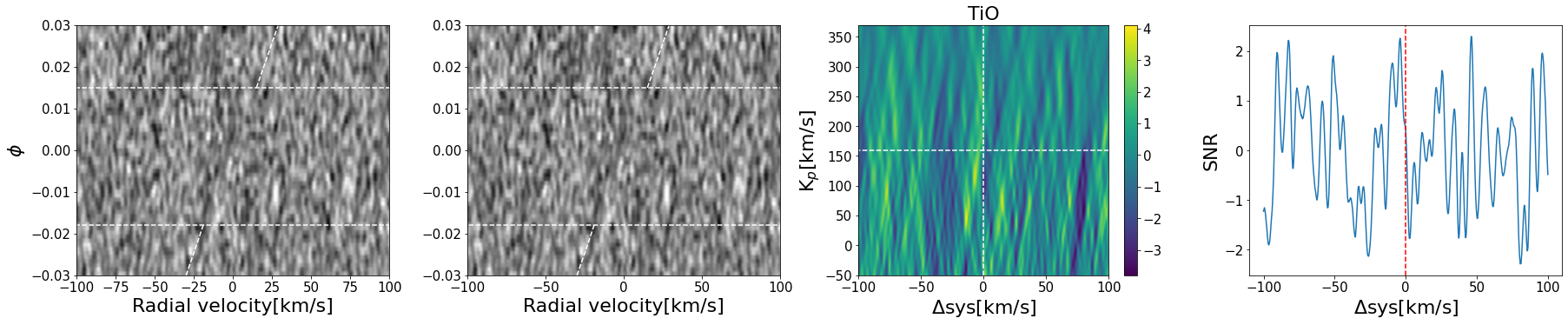}  
    \end{minipage}
    }
    \subfigure{
    \begin{minipage}[t]{1\linewidth}
    \centering
    \includegraphics[width=17cm]{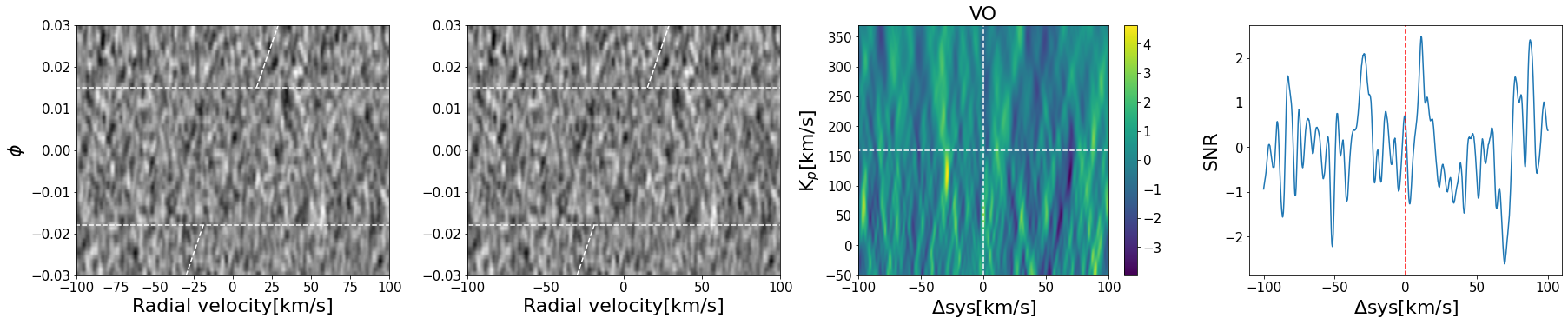}  
    \end{minipage}
    }

\caption{Same as Fig~\ref{ccf-result1}: but for \ion{Mg}{I}, \ion{Ti}{I}, \ion{V}{I}, \ion{Y}{I} TiO and VO.}
    \label{ccf-result2}
\end{figure*}  

\subsection{Cross-correlation function analysis}
\label{subsec:ccf}
\begin{figure} 
   \includegraphics[width=7cm, height=5cm]{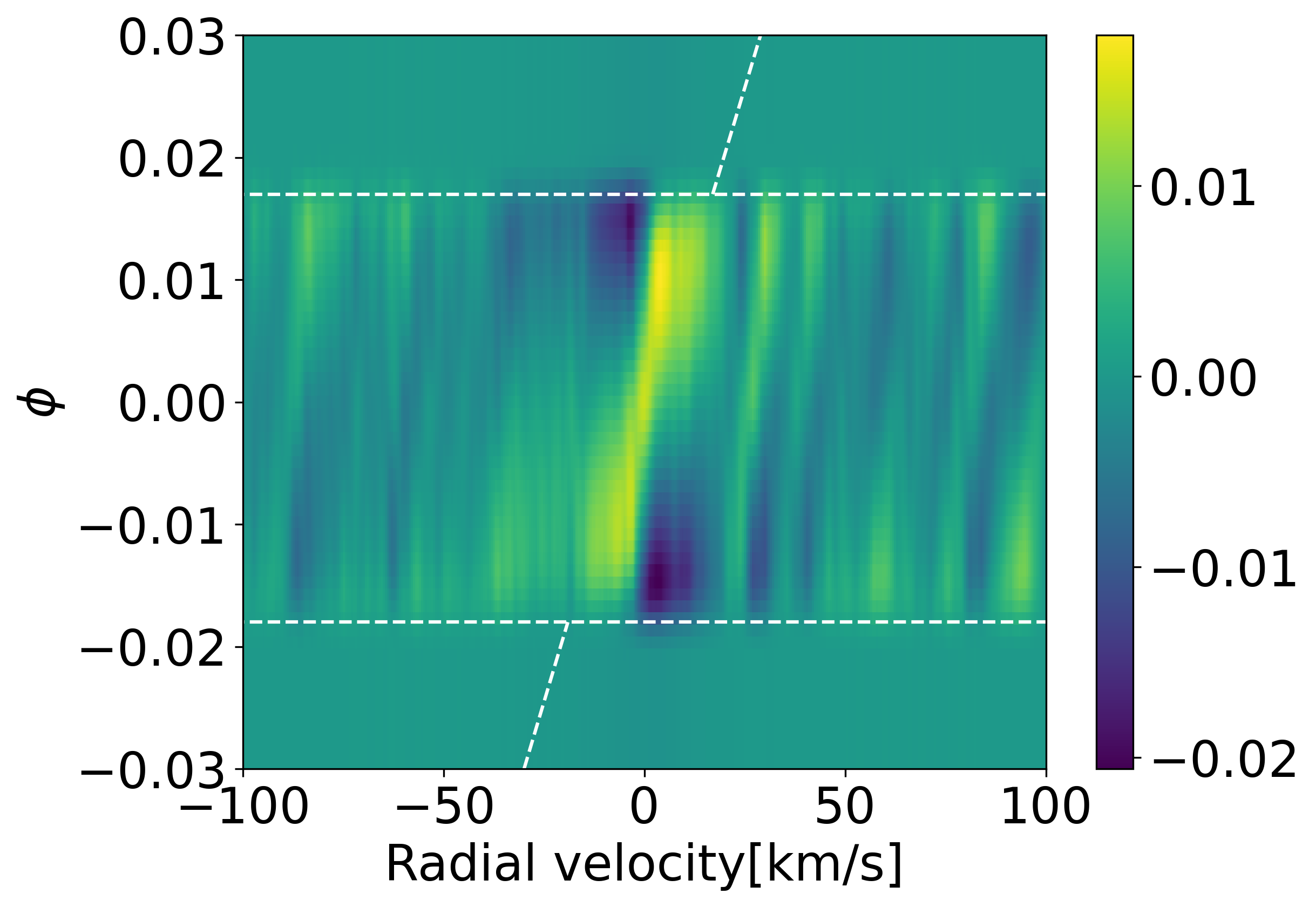}
   \centering
   \caption{The CCF map of model spectra generated by \texttt{SME} for \ion{Na}{I} on 2021 Feb 21, with only CLV and RM effects included, which is used to eliminate the influence of the variation of stellar line profile during transit.}
   \label{fig:RM+CLV ccf model}
  \end{figure}

In addition to the direct measurements of strong lines, we employ the CCF analysis to explore the presence of species or enhance their detection. The CCF method is to extract multiple spectral signals from a certain species by cross-correlating the template spectrum with only this species included with the observed transmission spectra. This technique is widely used for the high-resolution transmission spectroscopic study of exoplanets, which can fully utilize all the spectral signals from multiple lines of a certain species to maximize the detectability of that species. In this work, the species to be explored include \ion{Na}{I}, \ion{K}{I}, \ion{Ca}{I}, \ion{Cr}{I}, \ion{Fe}{I}, \ion{Li}{I}, \ion{Mg}{I}, \ion{Ti}{I}, \ion{V}{I}, \ion{Y}{I} and the molecules TiO and VO, and the spectrum template is computed by petitRADTRANS code \citep{Pmolliere_2019}. As input parameters, we assume the WASP-85Ab radius is 1.24$R_{\rm J}$ \citep{Mocnik_2016}, the planet mass is 1.265$M_{\rm J}$, and thus log\,$g_{\rm p}=3.329$ cgs. We apply an isothermal temperature of 1500\,K and assume a solar abundance for the determination of the volume mixing ratio (VMR) of various species. The template spectrum generated by petitRADTRANS is convolved to match the resolution of ESPRESSO. We set the radial  velocity at the range of $\pm100$\,km\,s$^{-1}$ with a step of 0.5\,km\,s$^{-1}$, and calculate CCF for each residual spectrum and model spectrum. The cross-correlation coefficients are calculated by the following formula:
\begin{equation} \label{ccf}
\centering
	c(v,t)= \frac{\sum_i^N x_i(t) T_i(v)}{\sum_i^N T_i(v)},
\end{equation}

where $T_i(v)$ is the template shifted to a radial velocity of $v$, $x_i(t)$ is the transmission spectrum at time $t$, $c(v,t)$ is a 2-dimensional matrix which is the function of $t$ and $v$. If the investigated species does exist, the signal will appear at the position of the estimated orbital velocity $K_{\rm p}$ and system velocity $V_{\rm sys}$\citep{Snellen_2010,Hoeijmakers_2019}.


The deformation of the stellar line profile caused by CLV and RM effects can also mask the planetary absorption signal in the cross-correlation maps which contains RM+CLV and planetary atmosphere signal together. It is necessary to correct these effects by modeling stellar lines at each phase. As described in Section~\ref{sec:data_analysis}, we used \texttt{SME} and the \texttt{VALD3} line list to compute theoretical stellar spectra that take into account the CLV and RM effects. The simulated stellar spectra are then cross-correlated with template spectra with a certain species, and the resulting 2D CCF map can be used as a proxy for the influence of the RM+CLV effects, and should be subtracted from the CCF map of the observed transmission spectra. The CCF map of the CLV+RM model CCF is shown in Fig.~\ref{fig:RM+CLV ccf model}.

The obtained 2D CCF map with CLV+RM effects uncorrected and corrected for \ion{Li}{I}, \ion{Na}{I}, \ion{Ca}{I}, \ion{K}{I}, \ion{Cr}{I}, and \ion{Fe}{I} are shown in the first and second panels of Fig.~\ref{ccf-result1}, where the inclined white dashes line indicates the expected track of planet signal. The third panels show the $K_{\rm p}-\Delta V_{\rm sys}$ map, in which the planet signal, if exists, should appear in the intersection of white dashed lines, i.e, with a velocity close to $V_{\rm sys}$  and an RV semi-amplitude $K_{\rm p}$ = 159.76 $\pm$\ 4.09 km s$^{-1}$. The fourth panels show the SNR of the corresponding species at the expected $K_{\rm p}$. From Fig.~\ref{ccf-result1}, only \ion{Li}{I} is clearly detected with a maximum SNR of $\sim4.5$, where the peak position corresponds to $K_{\rm p}$ of $\sim 134.8$\,$\pm$\ 88.43 km\,s$^{-1}$ and $\Delta V_{\rm sys} \sim$ -4.6 \,$\pm$\ 11.29 km\,s$^{-1}$. Some remaining features in the 2D maps may be associated with stellar activity~\citep{Stangret_2021} and propagated noise. The CCF maps and SNR plots for \ion{Mg}{I}, \ion{Ti}{I}, \ion{V}{I}, \ion{Y}{I} and TiO and VO are shown in Fig.~\ref{ccf-result2}, with no detection reported.


\section{Discussion and conclusion}
\label{sec:dis and conclusion}
We observed three transits of the hot Jupiter WASP-85Ab using the ultra-stable high-resolution spectrograph ESPRESSO. A total of 127 spectra were obtained, 64 spectra were taken during transits while the rest 63 spectra were taken out of transits. Telluric contamination was corrected for each spectrum, from which the master out-of-transit spectra were created and the transmission spectra were generated for the three transits. 

For the observed RV of the three transits, we performed a joint RM analysis, which can provide a good estimate of the origin of planets' angular momentum \citep{Ohta_2005}. We retrieved the projected obliquity value of ${\lambda = -16.155^{\circ}}^{+2.916}_{-2.879}$, suggesting that the hot Jupiter WASP-85Ab's orbit is almost aligned with its host star. The spectral signals induced by the CLV and RM effects were modeled and removed from the obtained transmission spectra and CCF map. The residual transmission spectra were used for the exploration of atomic and molecular lines that originate from WASP-85Ab's atmosphere, via direct inspection for strong lines or the CCF method for the species with multiple lines. The species we explored include \ion{H}{I},\ion{Li}{I},\ion{Na}{I}, \ion{Ca}{II}, \ion{K}{I}, \ion{Mg}{I}, \ion{Fe}{I}, \ion{Cr}{I}, TiO and VO. 

 \begin{figure} 
   \includegraphics[width=8cm, height=4cm]{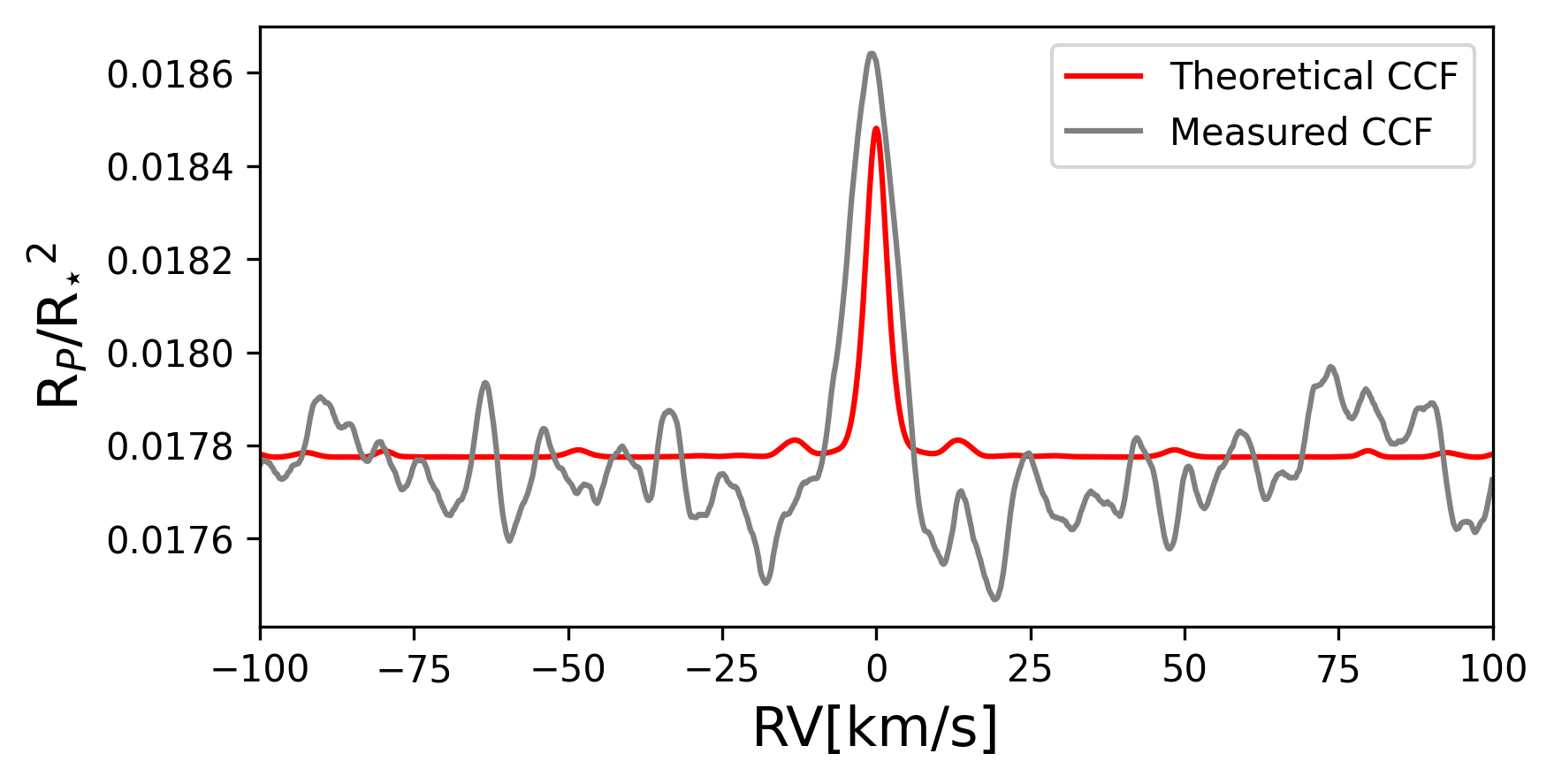}
   \centering
   \caption{Comparison between the measured CCF and theoretical CCF calculated by the template spectrum and observed data.}
   \label{upper_limit}
    \end{figure}

Among them, direct inspection revealed a $\sim10\sigma$ absorption signature at H$\alpha$ but it is not possible to confirm its origin as either stellar, planetary, or else. In the meanwhile, the \ion{Ca}{II}\,H \& K lines are also tentatively detected, with quite similar FWHM and consistent $R_{\lambda}$. The potential spectral signals detected from \ion{H}{I} and \ion{Ca}{II} may arise from the planetary atmosphere but we can not exclude yet the possibility of being induced by stellar activity. Another species that may be detected is \ion{Li}{I}, which has 4.5 $\sigma$ significance at the estimated $K_{\rm p}$ velocity in the $K_{\rm p}-\Delta V_{\rm sys}$ map. We note there is an offset of $21\pm4.3$\,km\,s$^{-1}$ between the estimated $K_{\rm p}$ and the retrieved $K_{\rm p}$. This discrepancy may be due to the large uncertainty of $K_{\rm p}$, given that the Li CCF signal is quite extended in the $K_{\rm p}$ direction. 

We note that there are still some structures visible in the 2D CCF maps, which may be the RM+CLV effect residuals that have not been removed completely, or be caused by the low SNRs in the stellar cores. We thus mask out the line cores with a width of 0.1\AA\ for the tentatively detected species to check whether there are signals left. As shown in Fig.~\ref{fig:masked atom line}, the absorption features of H${\alpha}$, \ion{Ca}{II}\,H\&K and \ion{Li}{I} are still visible, although slightly shallower. Therefore, the tentative detection of these three elements are unlikely to be caused by the RM+CLV effects or the low SNRs of the line cores.

The detection should not be affected by stellar activity, because there is no lithium line in the stellar spectrum. For the same reason, the continuum around the lithium lines possesses very high SNRs, resulting in a relatively strong lithium signal. Lithium was first reported in the atmosphere of WASP-127b by \citet{Chen_2018} with low-R spectroscopy but was not confirmed with high resolution in ~\citet{Allart_2020}.\,\citet{Borsa_2021} and \citet{Tabernero_2021} present the first detection of Li in the atmosphere of WASP-121b and WASP-76b respectively at high resolution. The detection of Li in an exoplanet atmosphere is not unexpected, as substellar objects with masses below $\sim$55\,$M_{\rm Jup}$ do not deplete this element during their lifetime \citep{Chabrier_2000,Baraffe_2015}. Such detection, if confirmed, is an important step, which can help understand of planet formation history and lithium depletion in planet-hosting stars \citep[e.g. ][]{Bouvier_2008,Chen_2018}.

Following the formalism of \citet{Wyttenbach_2015} and \citet{Allart_2017}, we estimate that the 5$\sigma$ upper limits of the excess absorption of the non-detected atoms and molecules, which are listed in Table~\ref{upper_limits}. Only lines with amplitudes higher than 10\,ppm are included in the template spectrum, and the corresponding number of lines employed are listed in Table~\ref{upper_limits} as well. Fig.~\ref{upper_limit} is an example showing how an upper limit is obtained by comparing the theoretical CCF using template spectrum with the measured CCF with the observed transmission spectra. 

\begin{table}
\centering
\caption[]{Number of lines used in the CCF calculations and upper limits derived for the investigated species.}
\label{upper_limits}
\begin{tabular}{ccc}
 \hline
 \hline
 \noalign{\smallskip}
Species & Number of lines& Detection limits  \\
\noalign{\smallskip}
\hline
\noalign{\smallskip}
\ion{Mg}{I} & 6 &527.3\,ppm\\
\noalign{\smallskip}
\ion{Li}{I} & 8 &323.9\,ppm\\
\noalign{\smallskip}
\ion{Fe}{I}&34&341.8\,ppm \\
\noalign{\smallskip}
\ion{Cr}{I}&45& 264.1\,ppm\\
\noalign{\smallskip}
\ion{K}{I}&53&364.2\,ppm\\
\noalign{\smallskip}
\ion{Y}{I}&99& 283.8\,ppm \\
\noalign{\smallskip}
\ion{V}{I}&154&301.6\,ppm\\
\noalign{\smallskip}
\ion{Ti}{I}&159&377.1\,ppm  \\
\noalign{\smallskip}
TiO & 7832 &27.6\,ppm \\
\noalign{\smallskip}
VO & 8280 &22.3\,ppm \\
\hline
\end{tabular}
\end{table}

It is somehow expected that the spectral features seen in the transmission spectra are very few and in general weak, given the relatively low $T_{\rm eq}$ and small $R_{\rm p}$ and thus small Transmission Spectroscopy Metric  ~\citep{Kempton_2018}, and that the star is active with recurring starspot reported which may induce relatively large noise in the derived spectra and 2D maps, although our measured S-index activity indicator of WASP-85A for three nights barely varies. In addition, the lack and weakness of atomic and molecular features can be due to several technical factors. The first reason is the low SNR of the observed stellar spectrum near the cores of the strong lines, where the photon counts are close to zero. Another factor is the overlap of CLV+RM effect and planet radial-velocity in cross-correlation residuals maps, making it difficult to disentangle the signal of an exoplanet atmosphere, as noted for example by \citet{Casasayas_Barris_2022}. More precise measurements of spin-orbit angle should in principle be able to better constrain the planet-occulating position, which will help eliminate the influence of the variation of stellar line profile during transit. 

We also note that the three nights data set does not comply with each other well. It seems the last night has the most significant spectral signals which should be related to the combined effect considering the lowest stellar activity indicated by S-index and lower impacts of WASP-85B showed by seeing value. More high-precision observations are necessary in order to further constrain and understand the atmosphere of WASP-85Ab.

\begin{acknowledgements}
We thank the anonymous reviewer for the constructive comments, thank Lauren Doyle for useful conversations and thank Jens Hoeijmakers for many useful suggestions. This research is the National Natural Science Foundation of China grants No.~11988101, 42075123, 42005098, 62127901, supported by the National Key R\&D Program of China No.~2019YFA0405102, the Strategic Priority Research Program of Chinese Academy of Sciences, Grant No.~XDA15072113, the China Manned Space Project with NO. CMS-CSST-2021-B12. MZ, YQS, QLOY are supported by the Chinese Academy of Sciences (CAS), through a grant to the CAS South America Center for Astronomy (CASSACA) in Santiago, Chile. HMC acknowledges support from a UKRI Future Leaders Fellowship (MR/S035214/1).
\end{acknowledgements}

%
%
\bibliographystyle{aa.bst}
\bibliography{cc.bib}



\onecolumn

\begin{appendix}
\section{Additional figures}
\label{sec:ap_indiv}

\begin{figure}[!ht] 
   \includegraphics[width=18cm]{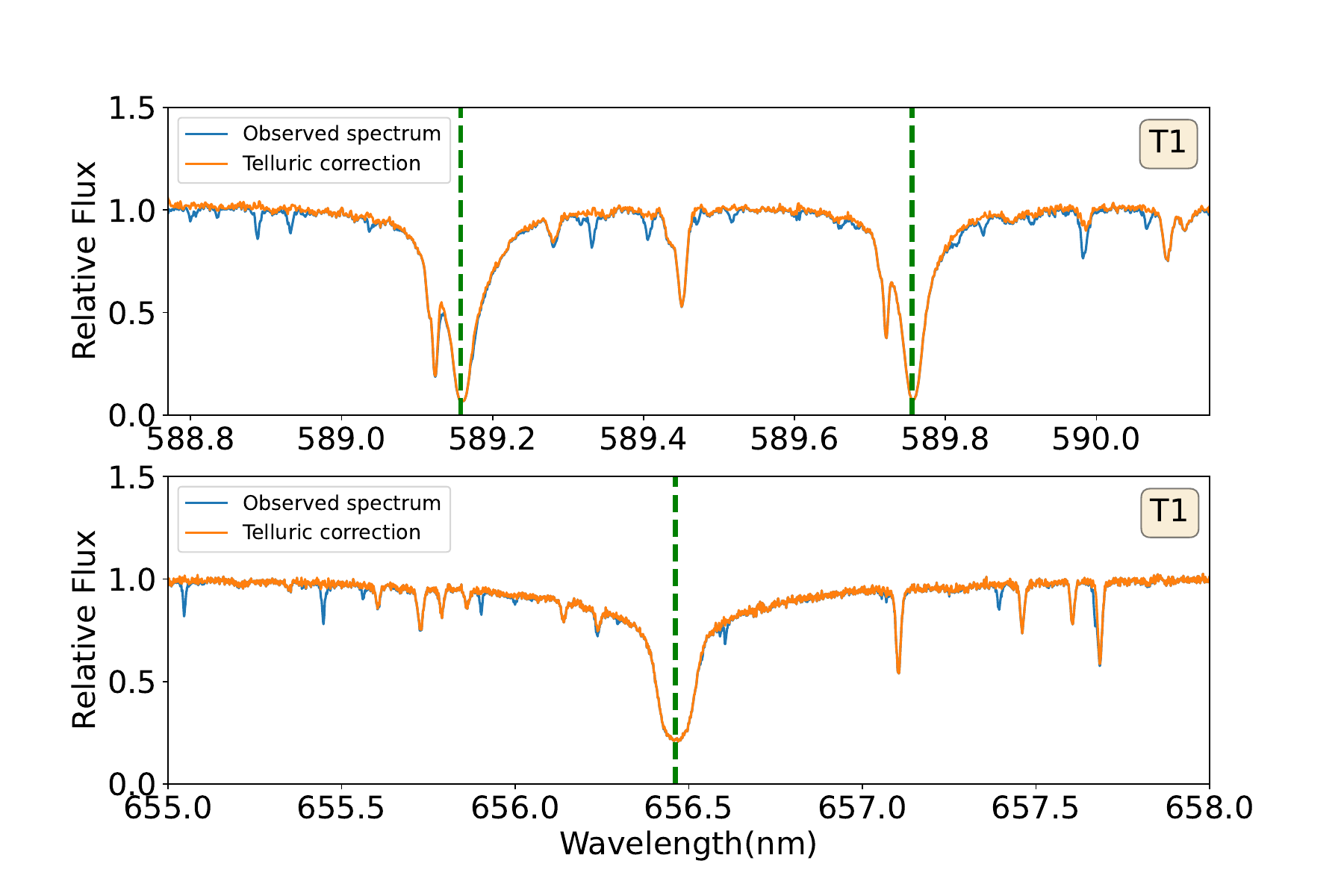}
   \centering
   \caption{The effect of telluric correction using Molecfit software in the observed spectrum of WASP-85Ab on the night of 2021 Feb 13 (T1) for single exposure. The observed spectrum is shown in blue line and the spectrum after telluric correction is shown in orange line. The dashed green vertical line represent the static position of atom spectral line at vacuum wavelength. Top panel: The telluric correction of spectral line around Na D1 $\&$ D2 lines. Bottom panel: The telluric correction of spectral line around H$\alpha$ lines.}
   \label{fig:Telluric_correction_effect}
    \end{figure}

\begin{figure}[!ht] 
   \includegraphics[width=13cm]{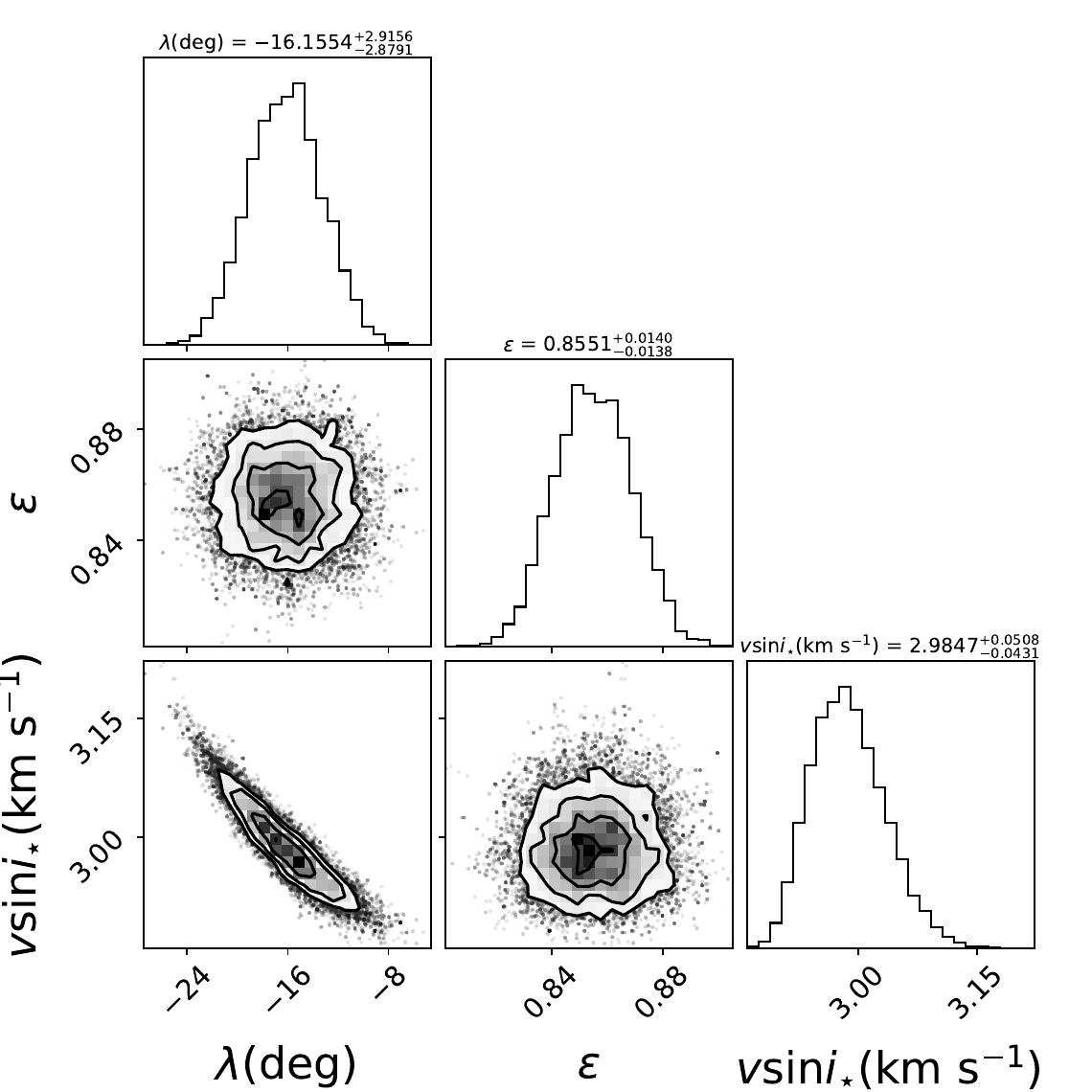}
   \centering
   \caption{Posterior distribution of the best-fit RM model using PyAstronomy and GP.}
   \label{fig:corner}
    \end{figure}

\begin{figure*}[h!]
  \includegraphics[width=17cm]{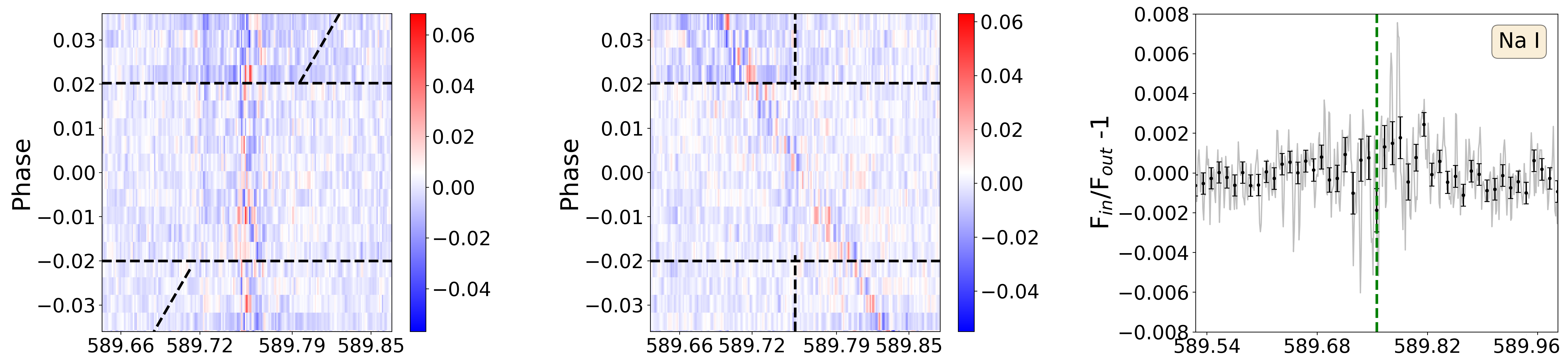}
  \includegraphics[width=17cm]{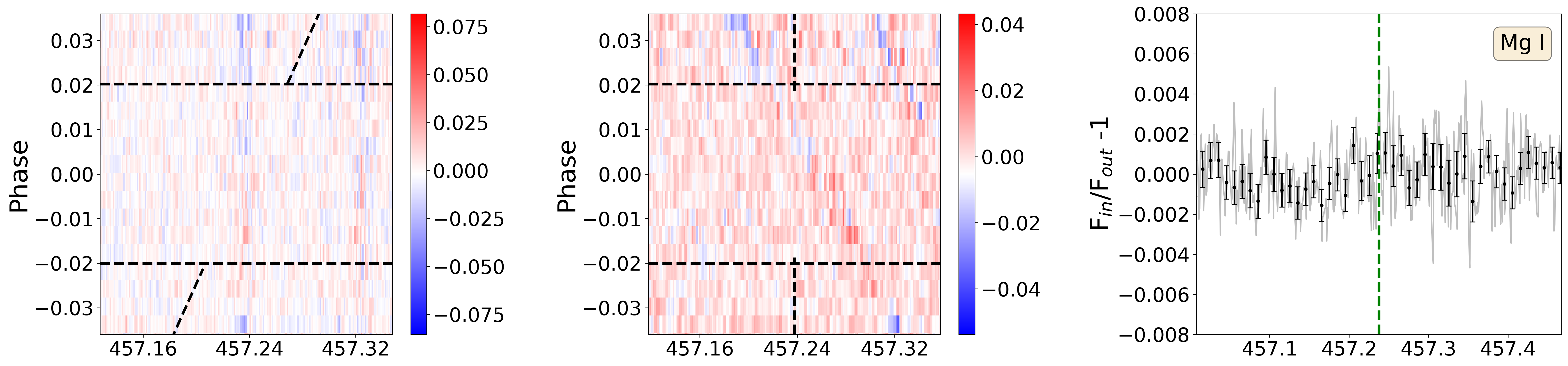}
  \includegraphics[width=17cm]{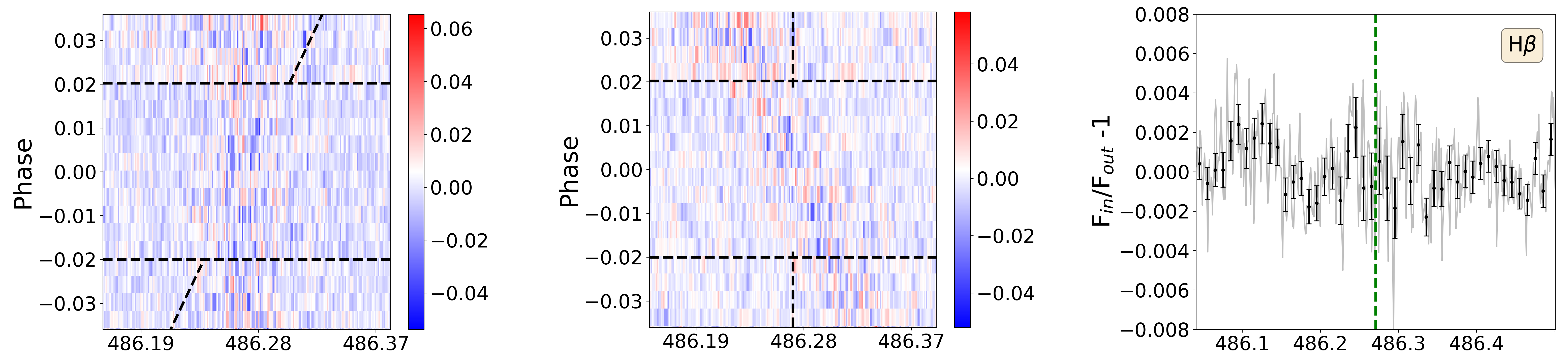}
  \includegraphics[width=17cm]{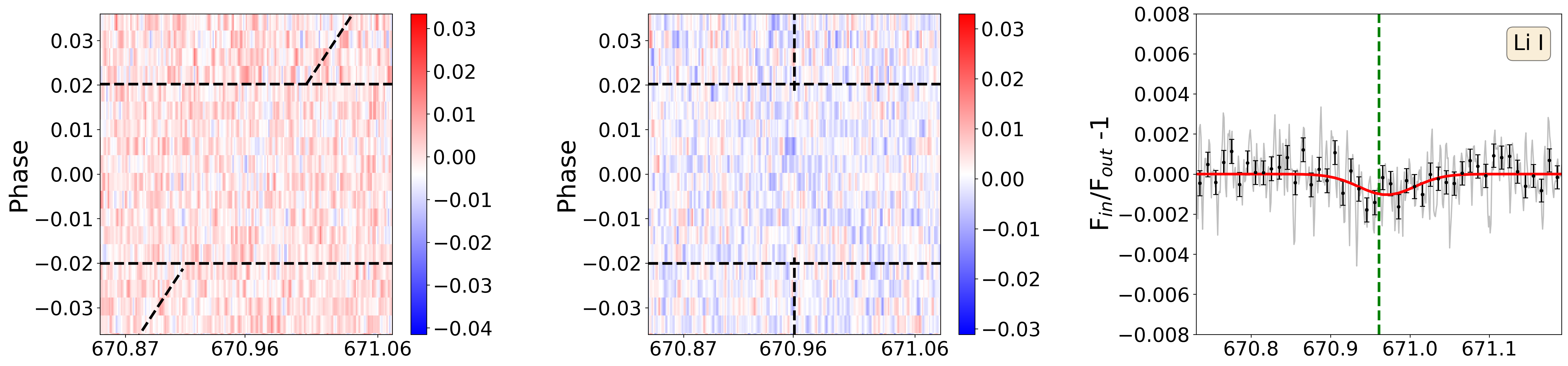}
  \includegraphics[width=17cm]{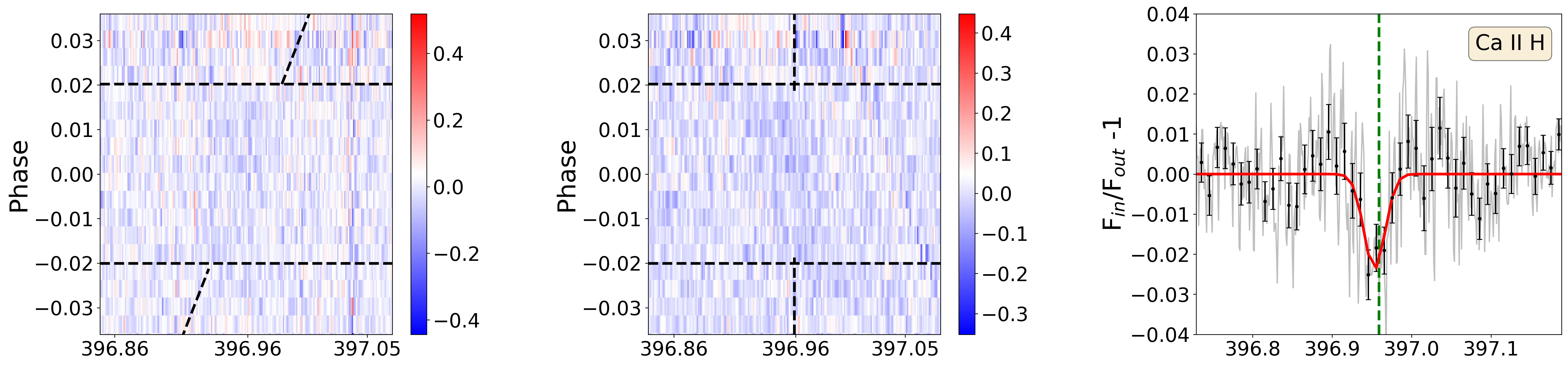}
  \includegraphics[width=17cm]{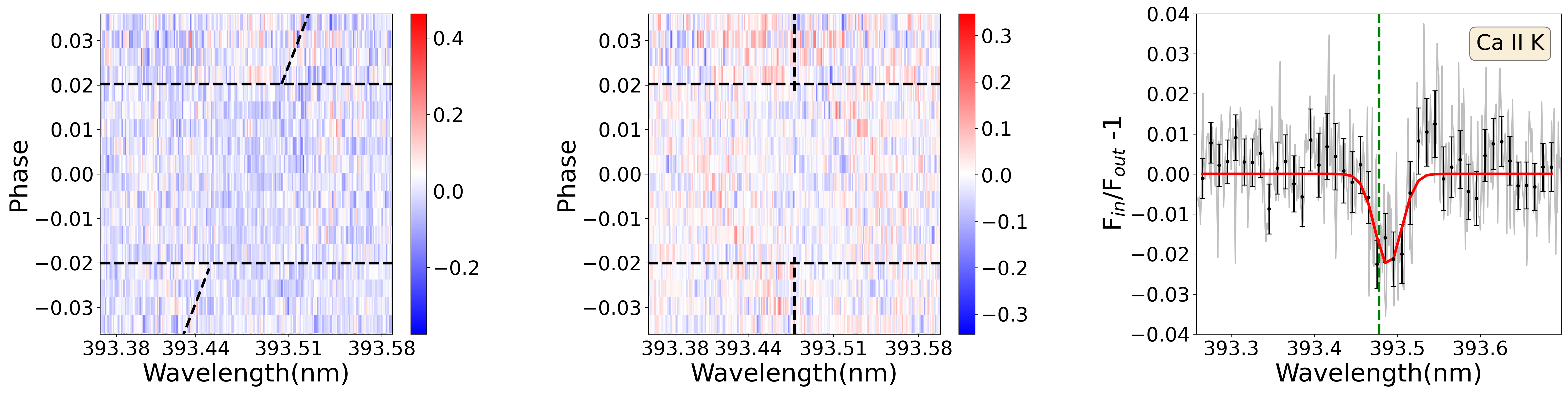}
  \caption{Same as Fig.~\ref{fig:Ha_line},but for \ion{Na}{I}, \ion{Mg}{I}, H${\beta}$, \ion{Li}{I}, \ion{Ca}{II} H and \ion{Ca}{II} K.}
  \label{fig:different atom line}
\end{figure*} 

\begin{figure*}[h!]
  \includegraphics[width=17cm]{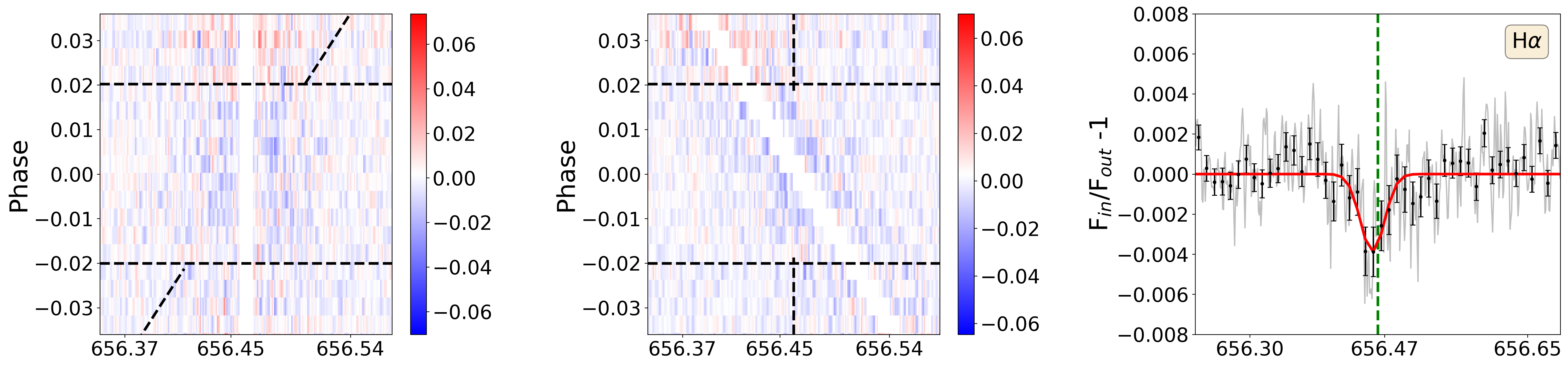}
  \includegraphics[width=17cm]{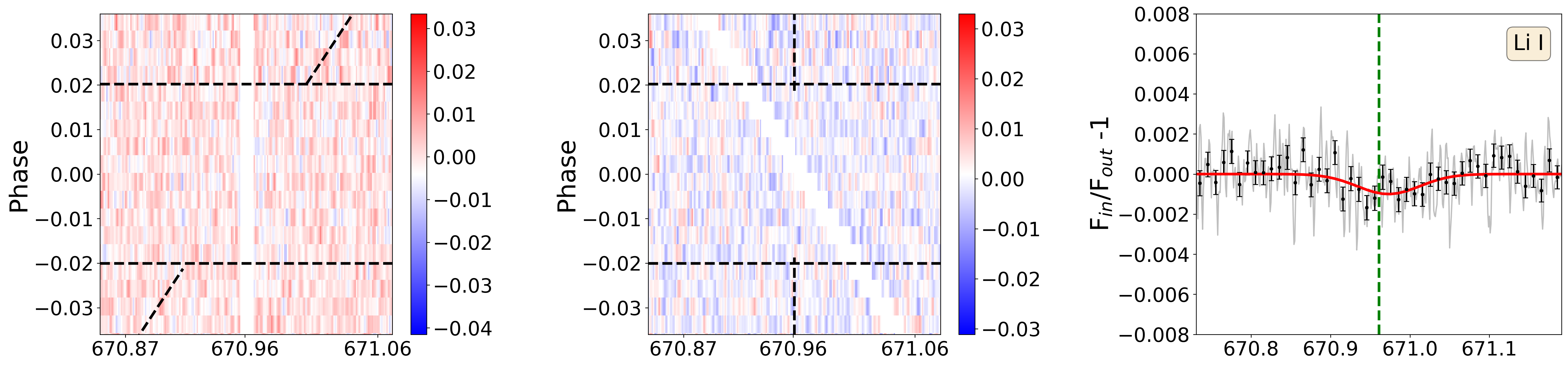}
  \includegraphics[width=17cm]{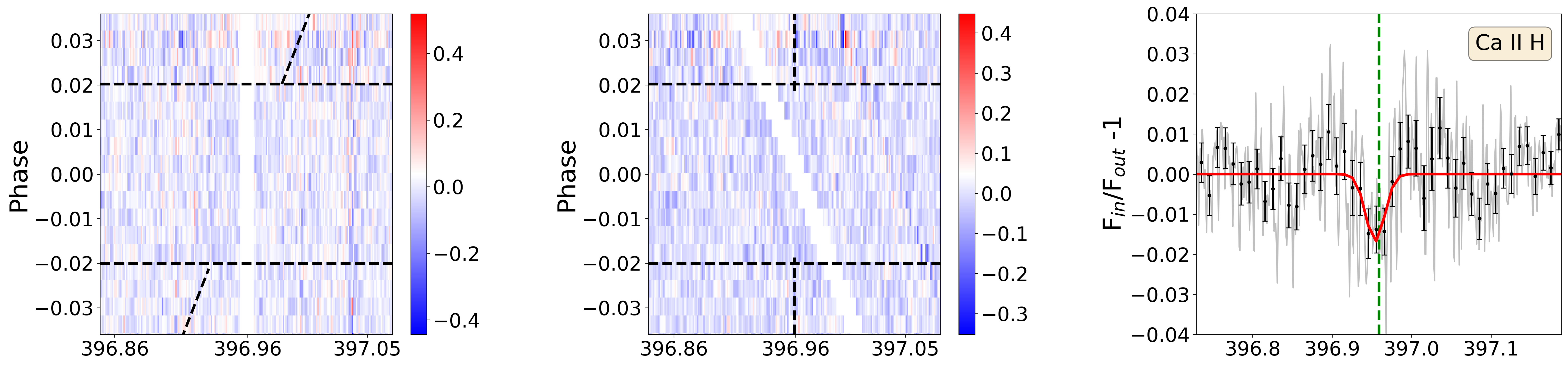}
  \includegraphics[width=17cm]{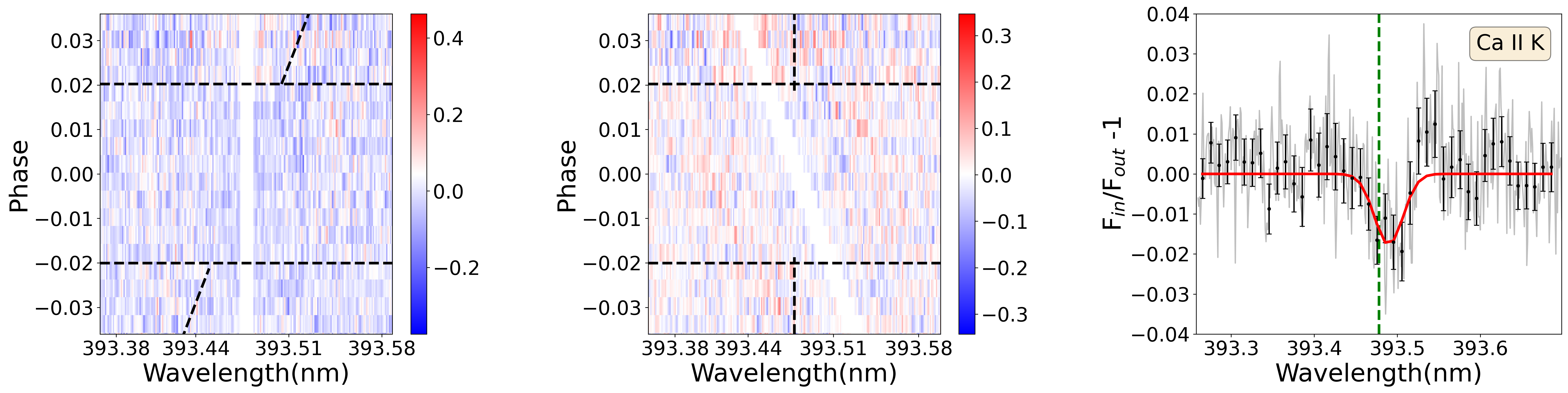}
  \caption{The masked phase-resolved 2D transmission spectra based on the combination of three nights for H$\alpha$, \ion{Li}{I}, \ion{Ca}{II} H and \ion{Ca}{II} K lines with RM+CLV correction applied in the left panel. The horizontal black-dashed lines indicate the beginning and end of the transit. The inclined black-dashed line presents the expected trace of signal from the exoplanet atmosphere. The middle panels show the 2D transmission spectra in the planet rest frame (PRF) assuming $K_{\rm p}$= 159.76\,km\,s$^{-1}$. In the right panels, the integrated transmission spectra in PRF with the CLV+RM effects corrected are shown in grey (original) and black (binned) and the best Gaussian fit are shown in red. The dashed green vertical line represents the static position of each line at vacuum wavelength.}
  \label{fig:masked atom line}
\end{figure*}

\begin{figure*}[htbp]
    \centering
    
    \subfigure{
    \begin{minipage}[t]{0.31\linewidth}
    \centering
    \includegraphics[width=5.5cm]{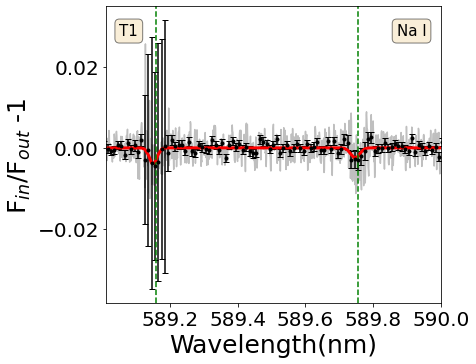}  
    \end{minipage}
    }
    \subfigure{
    \begin{minipage}[t]{0.31\linewidth}
    \centering
    \includegraphics[width=5.5cm]{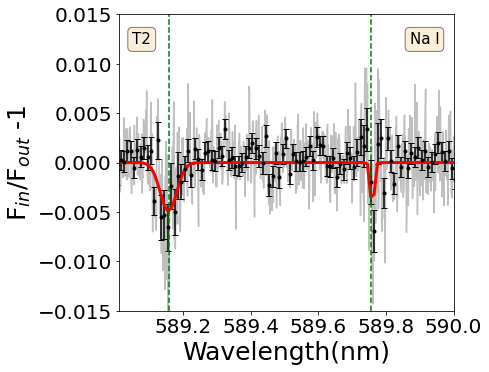}  
    \end{minipage}
    }
    \subfigure{
    \begin{minipage}[t]{0.31\linewidth}
    \centering
    \includegraphics[width=5.5cm]{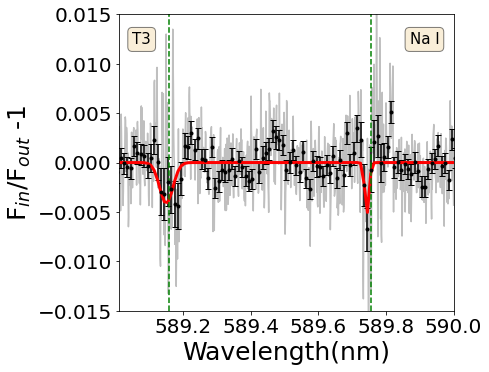}  
    \end{minipage}
    }
    \subfigure{
    \begin{minipage}[t]{0.31\linewidth}
    \centering
    \includegraphics[width=5.5cm]{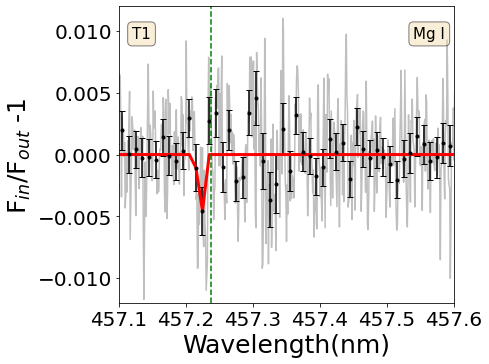}  
    \end{minipage}
    }
    \subfigure{
    \begin{minipage}[t]{0.31\linewidth}
    \centering
    \includegraphics[width=5.5cm]{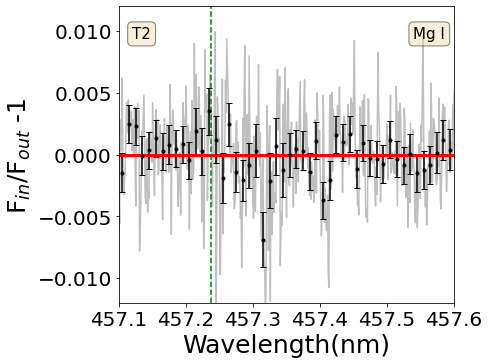}  
    \end{minipage}
    }
    \subfigure{
    \begin{minipage}[t]{0.31\linewidth}
    \centering
    \includegraphics[width=5.5cm]{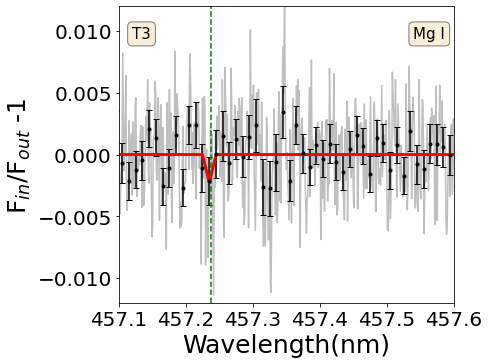}  
    \end{minipage}
    }     
    \subfigure{
    \begin{minipage}[t]{0.31\linewidth}
    \centering
    \includegraphics[width=5.5cm]{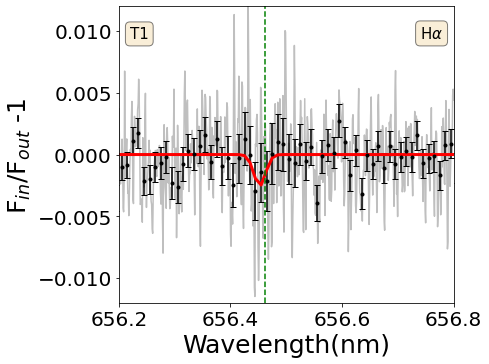}  
    \end{minipage}
    }
    \subfigure{
    \begin{minipage}[t]{0.31\linewidth}
    \centering
    \includegraphics[width=5.5cm]{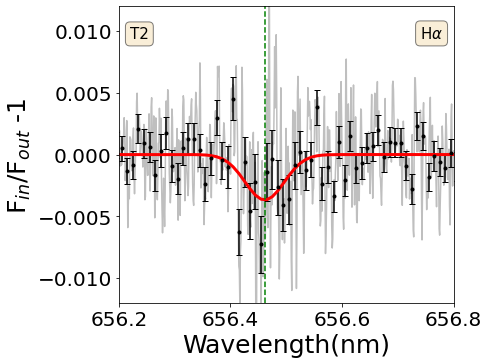}  
    \end{minipage}
    }
    \subfigure{
    \begin{minipage}[t]{0.31\linewidth}
    \centering
    \includegraphics[width=5.5cm]{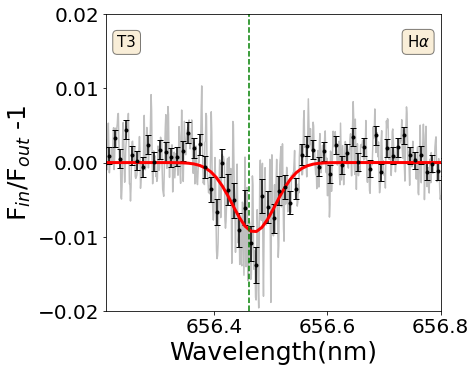}  
    \end{minipage}
    }
    \subfigure{
    \begin{minipage}[t]{0.31\linewidth}
    \centering
    \includegraphics[width=5.5cm]{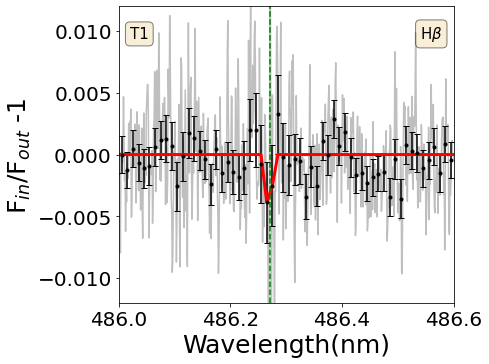}  
    \end{minipage}
    }
    \subfigure{
    \begin{minipage}[t]{0.31\linewidth}
    \centering
    \includegraphics[width=5.5cm]{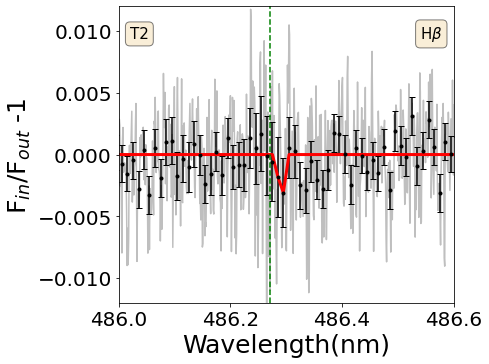}  
    \end{minipage}
    }
    \subfigure{
    \begin{minipage}[t]{0.31\linewidth}
    \centering
    \includegraphics[width=5.5cm]{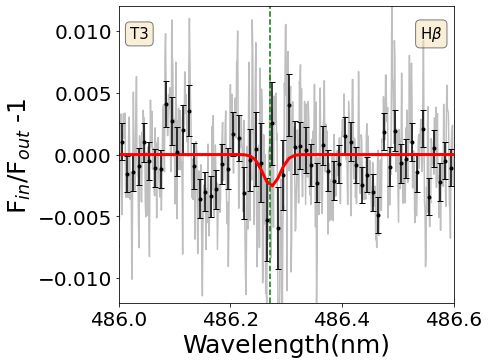}  
    \end{minipage}
    }  
    \caption{Transmission spectra of \ion{Na}{I}, \ion{Mg}{I}, H${\alpha}$, H${\beta}$ for each night. Left (T1), middle (T2), right (T3). The grey line represents the raw transmission spectra in the planet rest frame which has been corrected for CLV+RM effects, The black dotted line represents the binned spectra of  0.1 {\AA}   and the Gaussian fitting is shown in red line. The dashed green vertical line represent the static position of atom individual line at vacuum wavelength.}
    \label{fig:atom line-1}
\end{figure*}

\begin{figure*}[htbp]
    \centering
    \subfigure{
    \begin{minipage}[t]{0.31\linewidth}
    \centering
    \includegraphics[width=5.5cm]{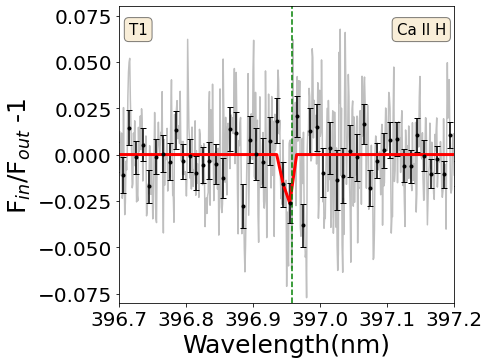}  
    \end{minipage}
    }
    \subfigure{
    \begin{minipage}[t]{0.31\linewidth}
    \centering
    \includegraphics[width=5.5cm]{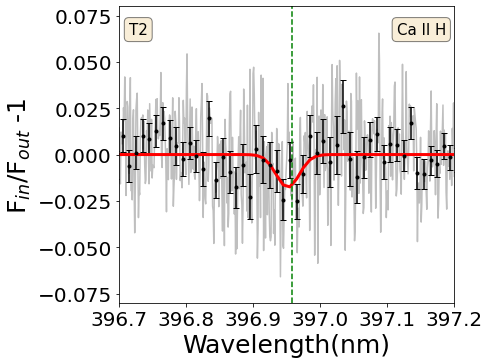}  
    \end{minipage}
    }
    \subfigure{
    \begin{minipage}[t]{0.31\linewidth}
    \centering
    \includegraphics[width=5.5cm]{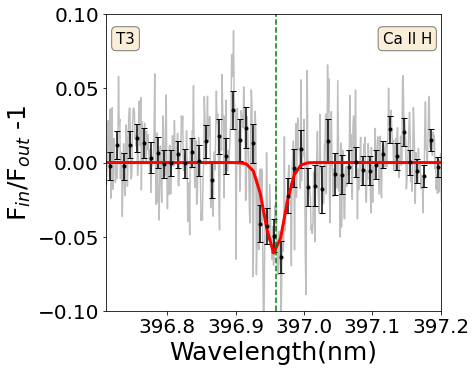}  
    \end{minipage}
    }
    \subfigure{
    \begin{minipage}[t]{0.31\linewidth}
    \centering
    \includegraphics[width=5.5cm]{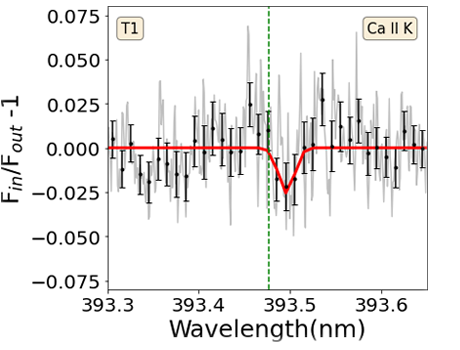}  
    \end{minipage}
    }
    \subfigure{
    \begin{minipage}[t]{0.31\linewidth}
    \centering
    \includegraphics[width=5.5cm]{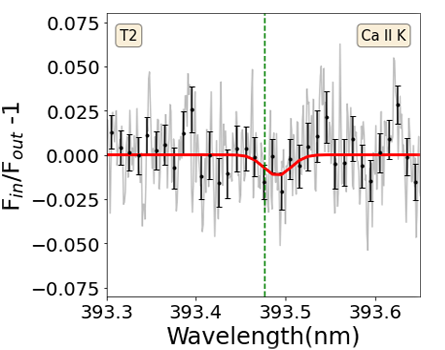}  
    \end{minipage}
    }
    \subfigure{
    \begin{minipage}[t]{0.31\linewidth}
    \centering
    \includegraphics[width=5.5cm]{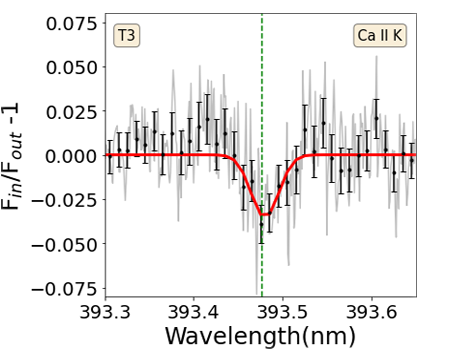}  
    \end{minipage}
    }  
   \subfigure{
    \begin{minipage}[t]{0.31\linewidth}
    \centering
    \includegraphics[width=5.5cm]{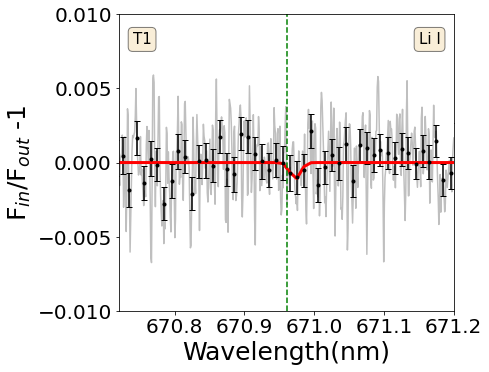}  
    \end{minipage}
    }
    \subfigure{
    \begin{minipage}[t]{0.31\linewidth}
    \centering
    \includegraphics[width=5.5cm]{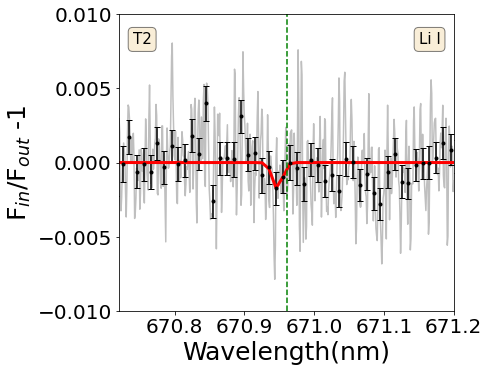}  
    \end{minipage}
    }
    \subfigure{
    \begin{minipage}[t]{0.31\linewidth}
    \centering
    \includegraphics[width=5.5cm]{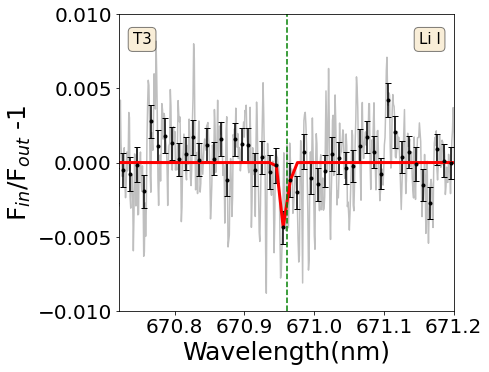}  
    \end{minipage}
    }
    
   \caption{Same as Fig~\ref{fig:atom line-1}, but for \ion{Ca}{II} H,\ion{Ca}{II} K and \ion{Li}{I}.}
    \label{fig:atom line-2}
 \end{figure*}
\begin{figure*}[!ht] 
   \includegraphics[width=18cm]{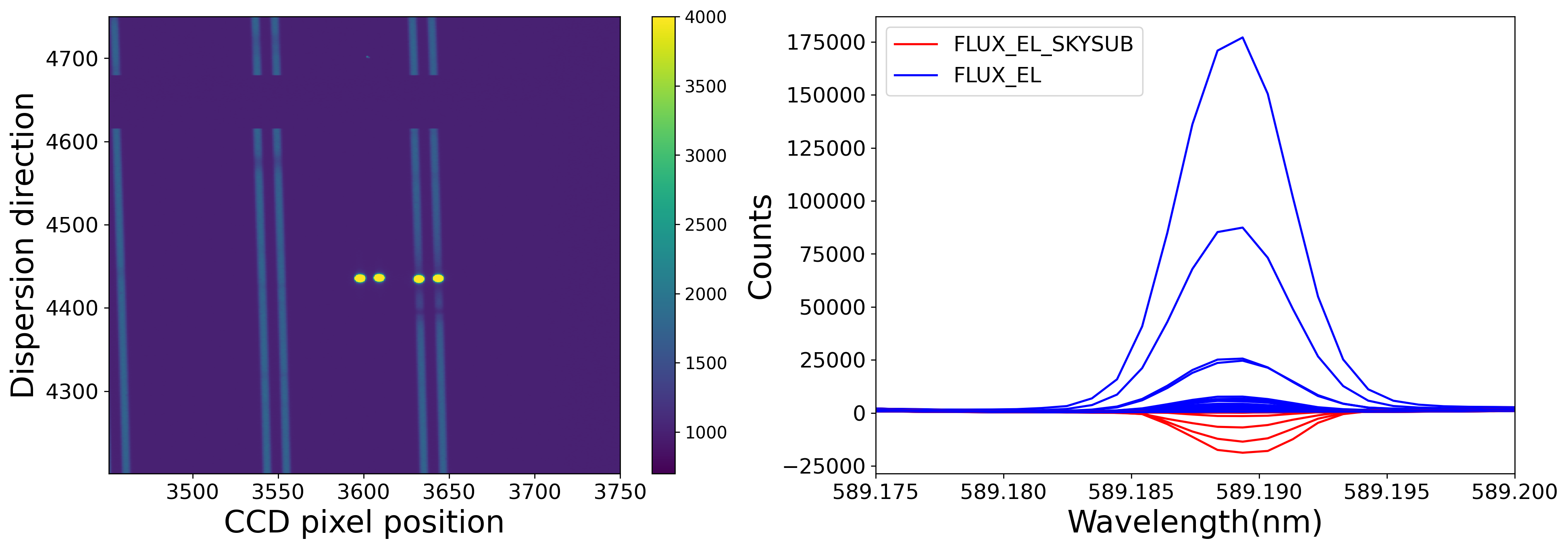}
   \centering
   \caption{This left panel shows that \ion{Na}{I} D2 absorption line in the raw data image of T1 is contaminated by two bright spikes which leads to the abnormal flux around \ion{Na}{I} D2 line. The right panel plots the abnormal counts of no sky-subtracted flux colored in blue and sky-subtracted flux shown in red. }
   \label{fig:Na_problem}
    \end{figure*}

\begin{figure*}[htbp]
  \centering
  \includegraphics[width=8cm]{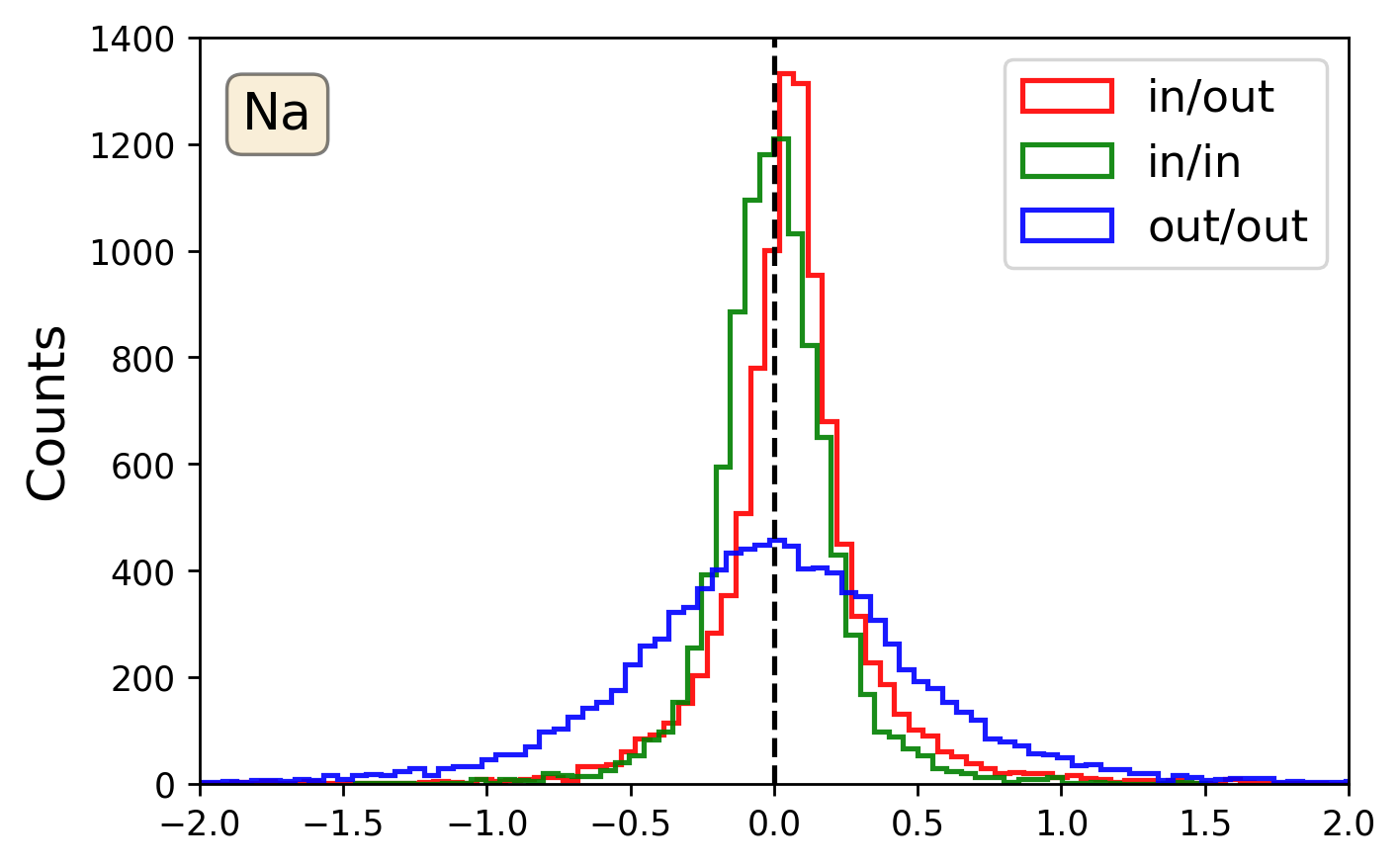}
  \includegraphics[width=8cm]{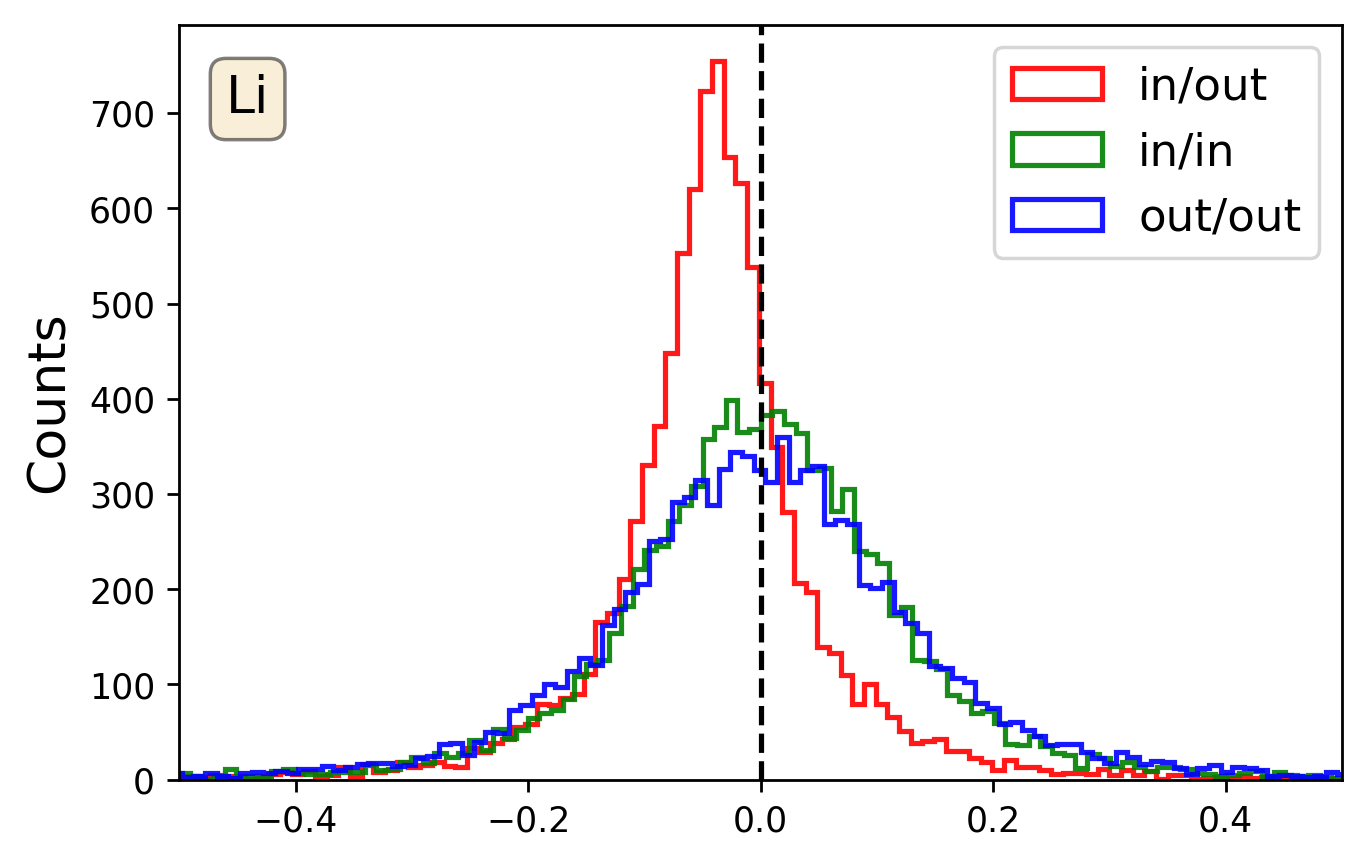}
  \includegraphics[width=8cm]{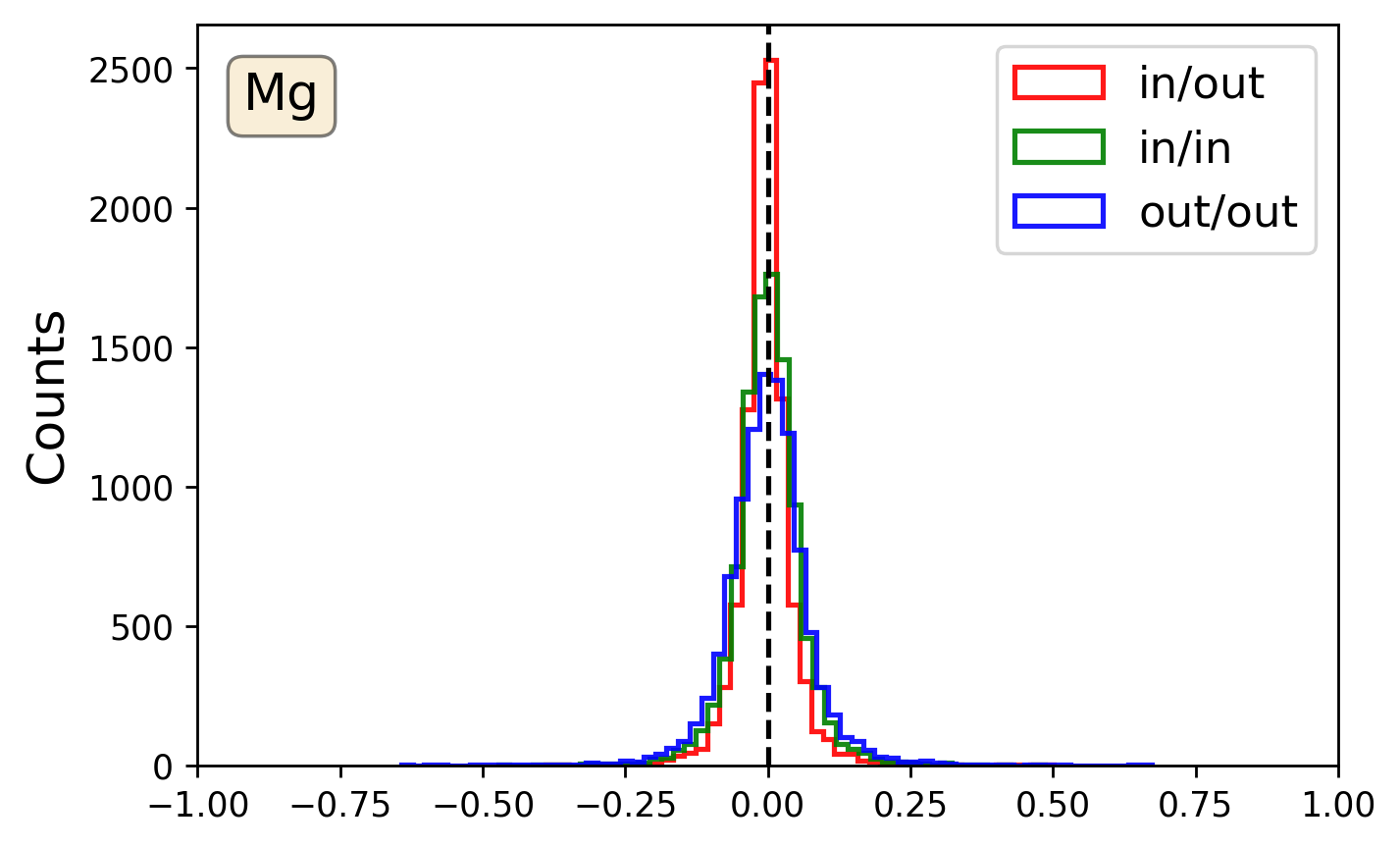}
  \includegraphics[width=8cm]{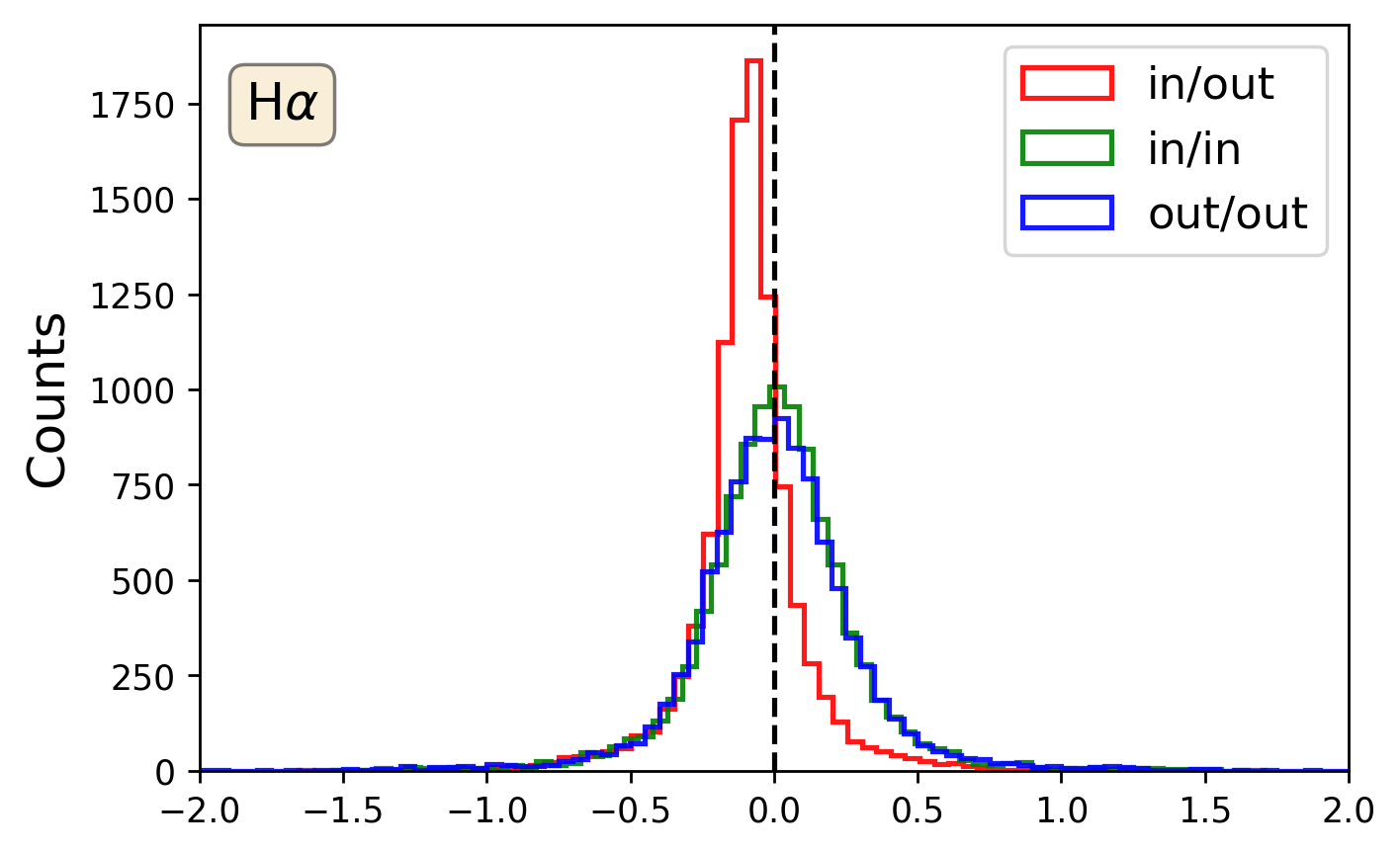}
  \includegraphics[width=8cm]{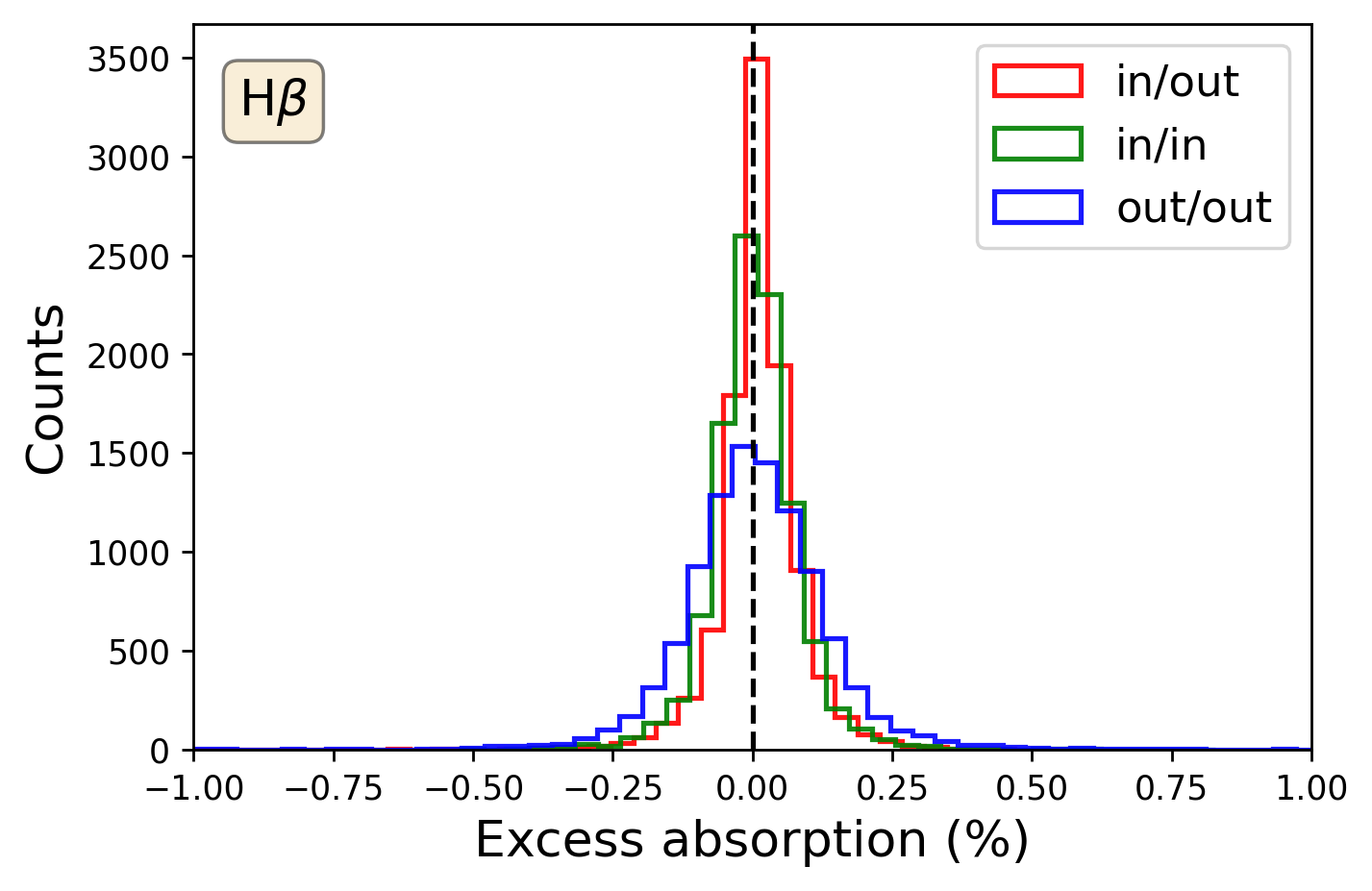}
  \includegraphics[width=8cm]{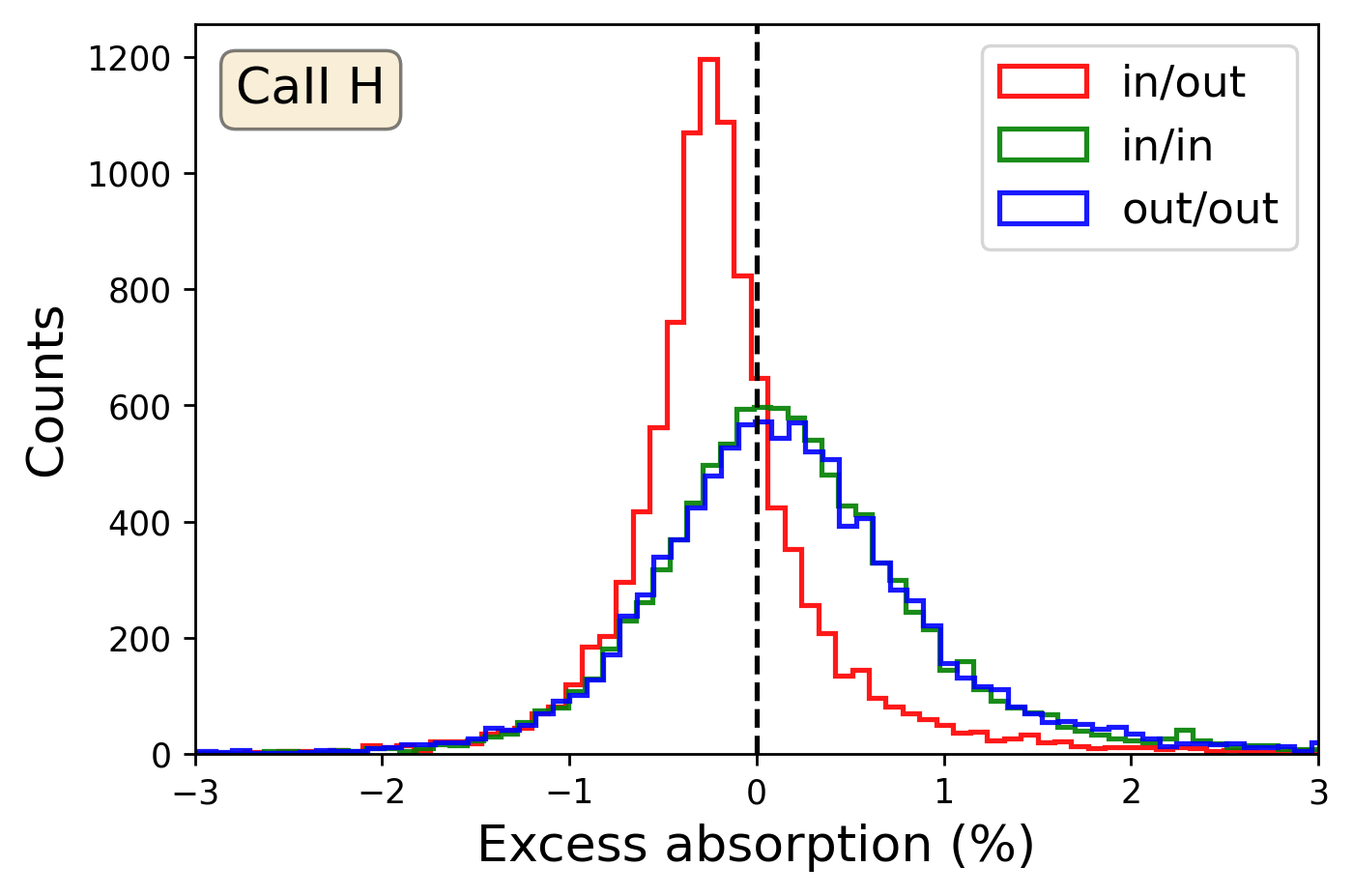}
  \includegraphics[width=8cm]{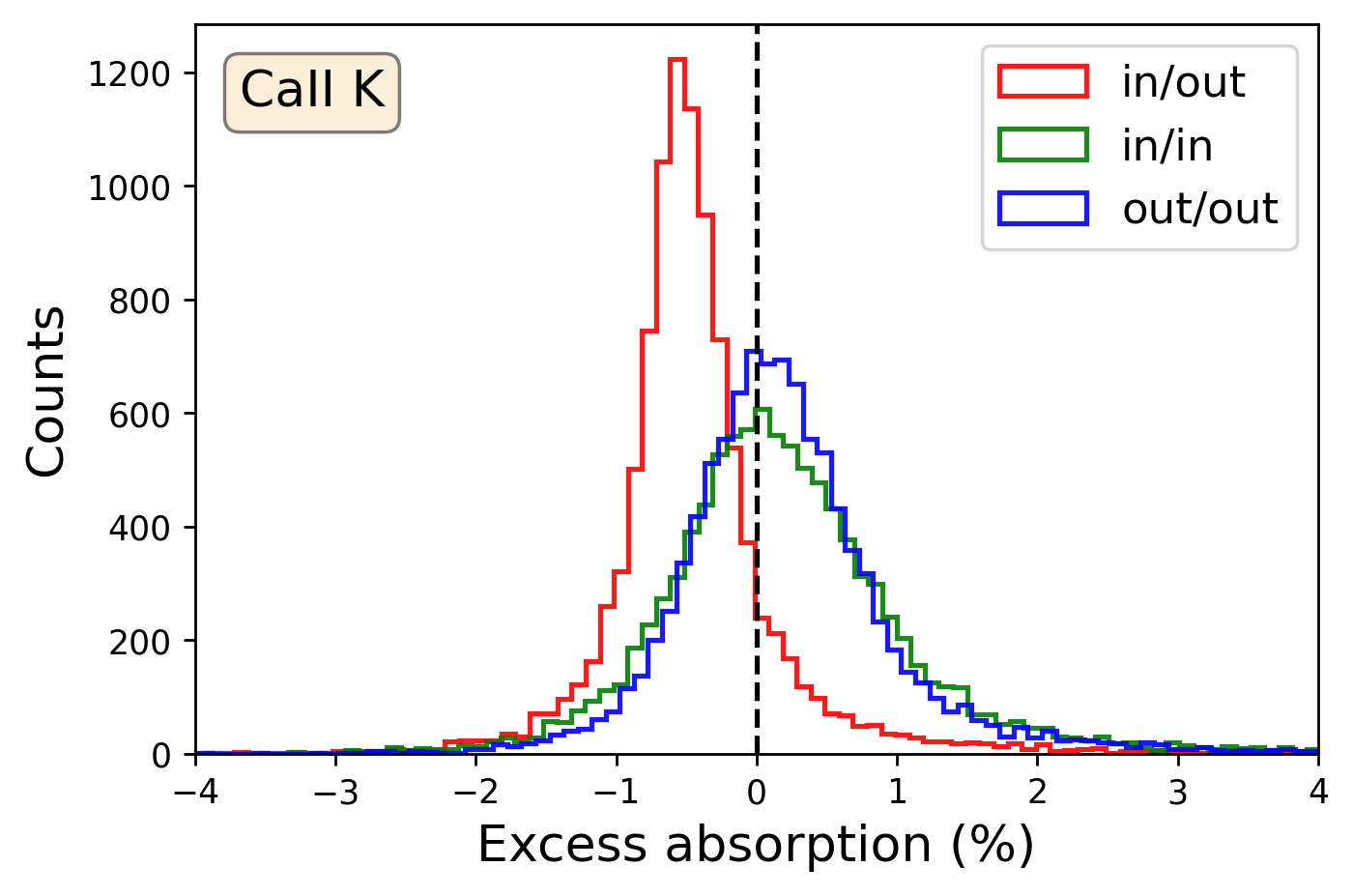}
  \caption{The results of EMC simulation for \ion{Na}{I}, \ion{Mg}{I}, \ion{Li}{I}, H${\alpha}$, H${\beta}$, \ion{Ca}{II} H and \ion{Ca}{II} K combined with three nights. These distributions for three EMC scenarios are shown in different colors, red is the in-out distribution; green is the in-in distribution; blue is the out-out distribution. The in-out distribution of \ion{Li}{I}, H${\alpha}$,  \ion{Ca}{II} H and \ion{Ca}{II} K exhibits excess absorption. }
  \label{fig:emc for atom}
\end{figure*} 

\end{appendix}

\end{document}